\documentclass[11pt]{article}

\usepackage[preprint]{acl}
\usepackage{times}
\usepackage{latexsym}
\usepackage[T1]{fontenc}
\usepackage[utf8]{inputenc}
\usepackage{adjustbox}
\usepackage{multicol}
\usepackage{amsmath,amsfonts,bm}

\def\eqref#1{equation~\ref{#1}}

\def\1{\bm{1}}

\DeclareMathAlphabet{\mathsfit}{\encodingdefault}{\sfdefault}{m}{sl}
\SetMathAlphabet{\mathsfit}{bold}{\encodingdefault}{\sfdefault}{bx}{n}

\newcommand{\KL}{D_{\mathrm{KL}}}

\usepackage{amsmath,amssymb}
\usepackage{booktabs}
\usepackage{algorithm}
\usepackage{algpseudocode}
\usepackage{tabularx}
\usepackage{array}
\usepackage{multirow}
\usepackage{float}
\usepackage{placeins}
\usepackage{needspace}
\usepackage{graphicx}
\usepackage{wrapfig}
\usepackage{pdflscape}
\usepackage{xcolor}
\usepackage{pgfplots}
\pgfplotsset{compat=1.18}
\usepackage{xspace}
\usepackage{listings}
\usepackage{microtype}
\usepackage{hyperref}
\usepackage{url}

\usepackage{todonotes}

\floatstyle{ruled}
\newfloat{algorithm}{tbp}{loa}
\floatname{algorithm}{Algorithm}

\title{ReSAT: Training LLMs Resistant against Refusal Ablation Attacks}

\author{
Yikun Zong\textsuperscript{1}\quad Yarin Gal\textsuperscript{2}\quad Lin Li\textsuperscript{2}\\[0.5em]
{\normalfont\textsuperscript{1}Department of Engineering, University of Cambridge, United Kingdom}\\
{\normalfont\textsuperscript{2}OATML, Department of Computer Science, University of Oxford, United Kingdom}\\[0.3em]{\normalfont\texttt{yz977@cam.ac.uk}\quad\texttt{lin.li@cs.ox.ac.uk}}
}

\hypersetup{pdfauthor={Yikun Zong, Yarin Gal, Lin Li}}

\begin{document}

\maketitle
\raggedbottom 

\begin{abstract}
Open-weight language models allow downstream users to modify internal computations as well as prompts, creating a distinct challenge for preserving learned safeguards. We investigate whether safety training protects refusal and safety against the removal of multiple activation directions, rather than just one. We introduce \emph{Refusal Subspace Ablation}, an attack that selects complementary directions and jointly removes their span throughout generation without updating model weights. Starting from one direction, it adds components outside the already selected span, choosing them by their effect under joint ablation. Our evaluations show that some current existing safety-trained models resist a selected single-direction attack yet remain vulnerable to joint ablation. To narrow this gap, we propose \emph{Refusal Subspace Ablation Training}: mixed supervised fine-tuning followed by safety-training rounds that supervise safe responses under joint ablation and refresh candidate directions between rounds. We evaluate ReSAT on Llama-3-8B, Mistral-7B, Qwen2.5-14B, against RSA and other refusal-ablation attacks, with jailbreak attacks and utility test. In key settings, ReSAT achieves lower attack success rates than most other baselines while preserving clean MMLU accuracy. By connecting adaptive attack evaluation with safety training, our work offers a practical path toward more robust safeguards for open-weight LLMs.
\end{abstract}

\section{Introduction}
\label{sec:introduction}

Open-weight language models enable local deployment, customization, and research, but also give downstream users access to the mechanisms responsible for refusal. Safeguarding these models therefore requires looking beyond adversarial prompts: users can fine-tune the weights or intervene directly in internal activations. Refusal-direction ablation makes this challenge concrete, suppressing refusal by removing an activation component without additional gradient-based training \citep{arditi2024refusal,wang2025multilingual}. The circulation of modified models highlights the practical relevance of this threat: four Qwen3.8-27B derivatives advertised as \emph{abliterated} or \emph{uncensored} recorded a combined 571,199 Hugging Face downloads in the rolling month reported on 12 September 2026.\footnote{Rolling monthly download counts from Hugging Face: \href{https://huggingface.co/Blackfrost-AI/Qwen3.8-27B-ABLITERATED-BF16}{Blackfrost-AI}; \href{https://huggingface.co/orcarouter/Qwen3.8-27B-Uncensored-FP8}{orcarouter}; \href{https://huggingface.co/huihui-ai/Huihui-Qwen3.8-27B-abliterated}{huihui-ai}; \href{https://huggingface.co/AEON-7/Qwen3.8-27B-AEON-ULTIMATE-UNCENSORED-BF16}{AEON-7}.}

Existing safety-training methods offer important defenses. MixAT combines continuous and discrete adversarial training \citep{dekany2025mixat}; ReFAT trains under dynamically refreshed refusal-feature ablation \citep{yu2024refat}; and DeepRefusal uses probabilistic ablation of an offline refusal direction \citep{xie2025deeprefusal}. Some resulting models retain high refusal rates under a selected single-direction attack. However, resistance to one direction does not establish resistance to the joint removal of complementary directions. Prior work, including Geometry \citep{wollschlager2025geometry} and SOM \citep{piras2026som}, has explored multi-directional refusal ablation, although their computational cost is relatively high. We ask: \emph{Do safety-trained models remain robust under adaptive joint ablation, and can training against cumulative interventions improve this robustness?}

We address the first question with \emph{Refusal Subspace Ablation}. Keeping the attacked model fixed, it iteratively selects additional complementary directions according to their effect under joint ablation. Each selected direction remains part of the intervention, and the entire accumulated subspace is removed throughout generation. Applied to safety-trained checkpoints, this evaluation exposes weaknesses missed by a single-direction test. For example, HarmBench ASR on Llama DeepRefusal rises from 0.00\% under single-direction ablation to 23.44\% under our RSA (Table~\ref{tab:main-three-family-results}).

This failure mode motivates \emph{Refusal Subspace Ablation Training}. After mixed supervised fine-tuning, we run SafeTrain rounds that supervise safe responses under joint removal of the accumulated refusal subspace. As training changes the model, we search for additional complementary directions between rounds, retaining the previously discovered span and expanding it when an eligible new direction is found. We evaluate the approach on Llama-3-8B, Mistral-7B, and Qwen2.5-14B. Our Llama H model achieves 11.20\% HarmBench ASR under rank-four RSA; its unablated full-test MMLU is 66.49\%, compared with 66.43\% for the original instruction-tuned baseline. These results support training against cumulative interventions, while keeping robustness to existing jailbreak and refusal ablation attacks. Hence, our main contributions are: 
\begin{itemize}
    \item \textbf{Efficient multi-directional refusal ablation.} RSA discovers complementary refusal directions and jointly removes their span, achieving attack effectiveness competitive with or even higher than Geometry and SOM while requiring substantially less measured search time.
    \item \textbf{Robust defense against refusal ablation.} Between SafeTrain rounds, ReSAT refreshes candidate directions while retaining the previously selected subspace and learns safe responses under cumulative removal, outperforming evaluated baselines in key settings and transferring to Geometry and SOM.
    \item \textbf{Complementary to jailbreak defense.} ReSAT complements existing adversarial training methods to achieve robustness against refusal ablation and jailbreak attacks jointly, while preserving general utility.
\end{itemize}

\section{Related Work}
\label{sec:related-work}

\noindent\textbf{Refusal directions and causal activation interventions.} Mechanistic studies of refusal identify low-dimensional activation features that separate harmful from harmless prompts and causally influence whether an instruction-tuned model refuses. A single residual-stream direction can be both necessary and sufficient for refusal across several model families: subtracting its component can induce harmful compliance, whereas adding it can elicit refusal on otherwise harmless inputs \citep{arditi2024refusal}. These findings establish activation intervention as a direct test of the internal mechanism, rather than merely another prompt-level jailbreak. They also motivate our central question: when safety training survives one known refusal direction, has it removed the vulnerability, or has the relevant computation migrated to a complementary subspace?

\noindent\textbf{Multi-directional refusal suppression.} Refusal need not be confined to one vector: Geometry \citep{wollschlager2025geometry} studies concept cones, representational independence, and joint interventions, while SOM \citep{piras2026som} develops multi-directional refusal suppression. Our contribution is an adaptive strong attack with low cost and more efficient search that selects complementary directions under joint ablation, which can be integrated into safety training.

\noindent\textbf{Adversarial safety training.} A complementary approach trains against input- or representation-space adversaries. Continuous Adversarial Training (CAT) replaces expensive discrete inner-loop search with continuous embedding perturbations and jointly rehearses benign utility data \citep{xhonneux2024cat}. Latent Adversarial Training (LAT) injects targeted perturbations into intermediate representations to suppress persistent harmful behavior \citep{sheshadri2024lat}. MixAT combines continuous and discrete adversarial examples to improve transfer across jailbreak families \citep{dekany2025mixat}. These methods expand the threat models encountered during training, but robustness to adversarial prompts or learned latent perturbations does not by itself establish robustness to causal removal of the feature that mediates refusal. We therefore evaluate them under interventions that repeatedly search and remove the currently exposed refusal component.

\noindent\textbf{Training against refusal-feature removal.} ReFAT directly simulates refusal-feature ablation during training and periodically refreshes the feature as the model changes \citep{yu2024refat}. DeepRefusal augments harmful examples with adversarial response prefixes and trains under probabilistic activation ablation, encouraging refusal to recover after the response has already entered a harmful trajectory \citep{xie2025deeprefusal}. Both approaches explicitly target refusal-feature removal and are consequently the closest baselines to our setting. The key distinction is the attack depth considered during optimization: a current or offline rank-one direction can become obsolete after the model reorganizes its computation. Between SafeTrain rounds, we extract fresh candidates from the updated checkpoint and search for a complementary direction while retaining the previously selected subspace. 

\section{Refusal Subspace Ablation}
\label{sec:attack}

Resistance to a single refusal direction does not establish resistance to removing several directions together. A model may still refuse after one component is removed, yet lose that behavior when complementary components are removed jointly. We therefore construct an attack that progressively expands the ablated subspace while retaining previously selected directions. The key is to evaluate each new direction \emph{together with the existing subspace}, rather than in isolation, while constraining changes on harmless prompts. 
We consider a white-box attacker who changes activations but keeps model weights fixed. The attack has two parts: a projection that removes a selected subspace, and an iterative search that decides which directions to include.

\subsection{From one direction to a joint subspace}
Let $h\in\mathbb R^d$ be a hidden activation and let the columns of $B\in\mathbb R^{d\times k}$ be $k$ orthonormal directions. The component $BB^\top h$ lies inside their span. Joint ablation subtracts this entire component, leaving only what lies outside:
\begin{equation}
 A_B(h)=h-BB^\top h,\qquad B^\top B=I_k.
 \label{eq:main-joint-ablation}
\end{equation}
One column recovers rank-one ablation; an empty $B$ gives the clean model. The same projection is applied at every decoder layer and token position, to the residual input and attention and MLP outputs, during both prompt processing and generation. Thus $B$ describes the components removed together, not a list of independent single-direction tests.

\subsection{Finding complementary directions by joint evaluation}
\paragraph{Step 1: extract candidates.}
Following prior work \citep{arditi2024refusal}, we compare mean activations on harmful and harmless prompts at candidate source layers and prompt positions. Their differences form a pool $\mathcal C^M$ for the fixed model $M$. These are candidate directions, not yet verified interventions. The source location specifies where a candidate is measured, not where it will be removed.

\paragraph{Step 2: remove already covered components.}
Starting from $B_0=\varnothing$, suppose $B_{k-1}$ contains the directions retained so far. Adding another vector within the same span would remove nothing new. We therefore subtract each candidate's projection onto that span, normalize its remaining component, and form a trial expanded basis:
\begin{equation}
\begin{aligned}
 r_k(c)&=(I_d-B_{k-1}B_{k-1}^\top)c,\\
 v_k(c)&=r_k(c)/\|r_k(c)\|_2,\\
 Q_k(c)&=\operatorname{Basis}([B_{k-1},v_k(c)]).
\end{aligned}
\label{eq:main-residual-expansion}
\end{equation}
Here $\operatorname{Basis}$ returns an orthonormal basis; negligible residuals are excluded. Orthogonalization avoids redundant directions, but does not establish that the new intervention suppresses refusal.

\paragraph{Step 3: evaluate the full intervention and retain one expansion.}
We temporarily ablate each complete trial basis $Q_k(c)$ and score the model on harmful validation prompts. For a selected set $R$ of refusal-probe tokens, $S_R$ averages the log-odds of generating a token in $R$ at the first assistant-output position; lower values favor refusal suppression. To limit nonspecific disruption, eligible candidates $\mathcal E_k$ must also satisfy a KL bound on the full first-token distribution for harmless prompts, relative to the clean model. We select
\begin{equation}
 c_k\in\arg\min_{c\in\mathcal E_k}S_R(Q_k(c);M),
 \qquad B_k=Q_k(c_k).
 \label{eq:main-joint-selection}
\end{equation}
The old directions remain active while the next expansion is tested. Raw clean candidates are cached for this fixed checkpoint; residuals and joint scores are recomputed at each step. Search stops at the requested rank or when no eligible candidate remains. The token score is only a selection proxy: generated refusal, utility, and semantic safety are evaluated separately. Appendix~\ref{app:attack-details} specifies the exact score, probes, eligibility filters, and seed protocols.

\section{ReSAT: \textbf{RE}fusal \textbf{S}ubspace \textbf{A}blation \textbf{T}raining}
\label{sec:training}

\begin{figure*}[t]
\centering
\includegraphics[width=\linewidth]{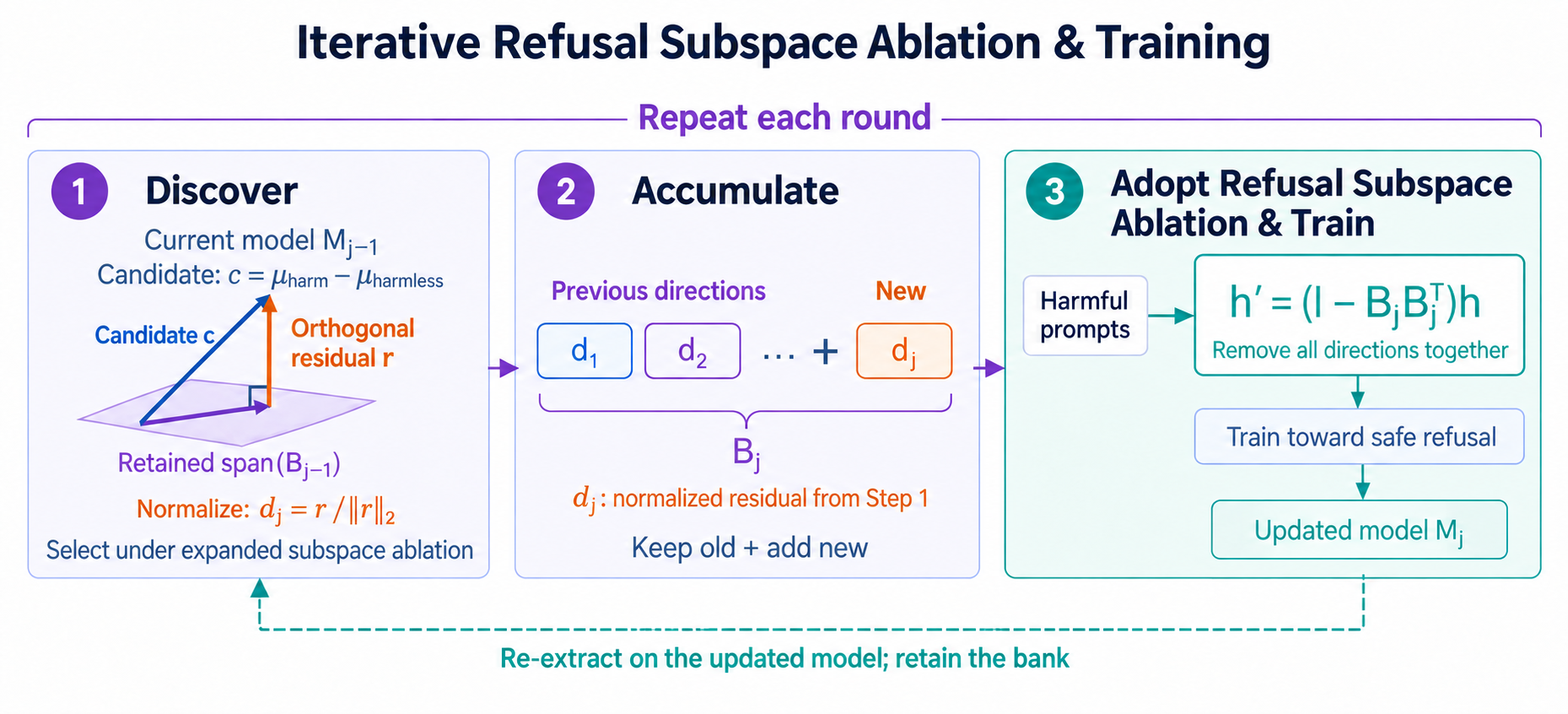}
\caption{\textbf{ReSAT training loop.} RSA expands the retained direction bank, and SafeTrain learns safe responses under joint ablation. Candidate directions are refreshed between rounds.}
\label{fig:resat-training-overview}
\end{figure*}

We train the model to produce safe responses while accumulated refusal subspace is jointly
removed. Our configuration uses mixed SFT initialization followed by iterative joint-ablation safety training (Figure~\ref{fig:resat-training-overview}). Only LoRA parameters $\theta$ are updated; pretrained weights remain frozen \citep{hu2021lora}.

\subsection{Iterative joint-ablation training}
\label{sec:iterative-safetrain}
Starting from the mixed-SFT adapter $\theta^{(0)}$ (Section~\ref{sec:mixed-sft}), initialize $B_0=\varnothing$. Before round $j$, freeze the current checkpoint $M_{\theta^{(j-1)}}$, extract fresh clean candidates, and apply the residualization and joint selection in Section~\ref{sec:attack} against the retained bank $B_{j-1}$. If an eligible new direction $d_j$ is found, update
\begin{equation}
\begin{aligned}
 B_j&=\operatorname{Basis}([B_{j-1},d_j]),\\
 \theta^{(j)}&=\operatorname{Optimize}
 \bigl(\theta^{(j-1)},\mathcal L_j(\,\cdot\,;B_j)\bigr).
\end{aligned}
\label{eq:main-training-loop}
\end{equation}
The complete $B_j$ stays fixed within the round, while gradients pass through its projection to the LoRA parameters. If no eligible new direction is found, the original H procedure stops and retains the last completed round. Explicit fixed-bank variants instead continue with $B_j=B_{j-1}$; their rank does not increase. 


\paragraph{A shared response loss.}
All training conditions supervise the assistant's target response. For target $y$, intervention $B$, and detached input perturbation $\bar\delta$, define
\begin{equation}
\begin{split}
 &\ell_{\rm resp}(\theta;x,y;B,\bar\delta)\\
 &=-\frac{1}{|y|}\sum_{t=1}^{|y|}
 \log p_{\theta,B,\bar\delta}(y_t\mid x,y_{<t}).
\end{split}
 \label{eq:main-response-loss}
\end{equation}
Here $x$ is the prompt, $y_t$ the target token at position $t$, $y_{<t}$ its target prefix, and $|y|$ the supervised response length. $p_{\theta,B,\bar\delta}$ is the model's conditional distribution under ablation $B$ and input perturbation $\bar\delta$; outer optimization treats $\bar\delta$ as constant. We average over response tokens, then examples, excluding user and padding tokens. $B=\varnothing$ and $\bar\delta=0$ denote no ablation and no input attack, respectively. The training objective has five terms:
\begin{equation}
\begin{aligned}
 \mathcal L_j={}&\lambda_{\rm clean}\mathcal L_{\rm clean}
 +\lambda_{\rm joint}\mathcal L_{\rm joint}
 +\lambda_{\rm cont}\mathcal L_{\rm cont}\\
 &+\lambda_{\rm pap}\mathcal L_{\rm pap+cont}
 +\lambda_{\rm help}\mathcal L_{\rm help}.
\end{aligned}
\label{eq:main-training-objective}
\end{equation}
Each term is an average of Equation~\ref{eq:main-response-loss} over its presentation distribution; the nonnegative $\lambda$ coefficients retain the original sampling shares and sum to one.

\paragraph{Clean refusal and joint-ablation refusal.}
For harmful prompts $x$ with safe refusal targets $y_{\rm ref}$, $\mathcal L_{\rm clean}$ averages $\ell_{\rm resp}(\theta;x,y_{\rm ref};\varnothing,0)$, maintaining refusal without intervention. $\mathcal L_{\rm joint}$ instead averages $\ell_{\rm resp}(\theta;x,y_{\rm ref};B_j,0)$, directly teaching refusal when the \emph{entire accumulated subspace} is removed. The two terms use the same normalized harmful-prompt distribution and targets, but different forward computations.

\paragraph{Input attacks combined with joint ablation.}
$\mathcal L_{\rm cont}$ averages $\ell_{\rm resp}(\theta;x,y_{\rm ref};B_j,\bar\delta_x)$ after continuous perturbation search on user-token embeddings. $\mathcal L_{\rm pap+cont}$ first samples a persuasive adversarial prompt (PAP) rewrite $\widetilde x=P(x;\xi)$, where $P$ is the rewriting procedure and $\xi$ its sampling randomness, then averages $\ell_{\rm resp}(\theta;\widetilde x,y_{\rm ref};B_j,\bar\delta_{\widetilde x})$ over examples and rewrites. Both branches keep $B_j$ active during inner attack search and outer training. The inner search increases the likelihood of an unsafe target prefix; the outer update always supervises the safe refusal. Perturbations are aligned to the outer input and detached: neither the attack optimizer nor direction selection is differentiated through.

\paragraph{Helpful responses on benign inputs.}
We group utility and hard-benign supervision under $\mathcal L_{\rm help}$ while retaining their distinct roles. For each pool $a\in\{\mathrm{util},\mathrm{hb}\}$, define
\begin{equation}
 \mathcal L_a(\theta)=\mathbb E_{(x,y)\sim\mathcal D_a}
 \ell_{\rm resp}(\theta;x,y;\varnothing,0).
 \label{eq:main-help-components}
\end{equation}
Here $\mathcal D_{\rm util}$ contains general-task examples with helpful targets, so $\mathcal L_{\rm util}$ preserves general capability. $\mathcal D_{\rm hb}$ contains harmless requests prone to over-refusal, also paired with helpful rather than refusal targets; $\mathcal L_{\rm hb}$ teaches the model to answer these requests normally. Each loss averages the response-normalized NLL within its own pool, without ablation or input perturbation. The grouped contribution is exactly
\begin{equation}
\begin{aligned}
 \lambda_{\rm help}\mathcal L_{\rm help}
 &=\lambda_{\rm util}\mathcal L_{\rm util}
 +\lambda_{\rm hb}\mathcal L_{\rm hb},\\
 \lambda_{\rm help}&=\lambda_{\rm util}+\lambda_{\rm hb}.
\end{aligned}
\label{eq:main-help-mixture}
\end{equation}
For SafeTrain, $\lambda_{\rm util}=0.35$ and $\lambda_{\rm hb}=0.20$ give $\lambda_{\rm help}=0.55$. This grouping preserves the six-condition recipe and its relative weights. There is no extra KL penalty in the outer loss: KL constrains direction selection only. The sampling shares, round count, branch definitions, token alignment, and optimization mechanics are in Appendix~\ref{app:training-details}.

\subsection{Mixed SFT initialization}
\label{sec:mixed-sft}
Mixed SFT aims to concentrate refusal around an identifiable direction while obscuring the base model's original direction, making attacks that reuse it less effective. Starting from the instruction-tuned checkpoint with a fresh LoRA adapter, we combine safe-refusal targets for harmful inputs with helpful answers for utility and hard-benign prompts. Harmful branches apply continuous attacks to both original and PAP-rewritten prompts \citep{zeng2024pap}; no direction ablation is used during SFT. The resulting $\theta^{(0)}$ initializes SafeTrain, which continues the same adapter under cumulative direction removal. Appendix~\ref{app:ablation-study-details} gives the SFT composition; Section~\ref{sec:ablation-study} reports the ablations.

\section{Results}
\label{sec:comparison}

\begingroup
\begin{table*}[t]
\caption{Refusal-ablation performance of base models and their safety-trained variants (\%). Protocols and cohorts are specified in Appendix~\ref{app:main-table-protocol}. Refusal Rate evaluation is in Appendix~\ref{app:lexical-refusal-rates}.} 
\label{tab:main-three-family-results}
\label{tab:main-ablation-results}
\centering
\normalsize
\setlength{\tabcolsep}{2pt}
\renewcommand{\arraystretch}{1.12}
\begin{adjustbox}{max width=\linewidth}
\begin{tabular}{@{}l*{5}{c}!{\hspace{3pt}\vrule width 0.4pt\hspace{3pt}}*{5}{c}@{}}
\toprule& 
\multicolumn{5}{c}{ASR $\downarrow$} & 
\multicolumn{5}{c}{MMLU $\uparrow$} \\

\cmidrule(lr){2-6}
\cmidrule(lr){7-11}Model & Clean & Single & Geometry & SOM & \shortstack{RSA\\(Ours)} & Clean & Single & Geometry & SOM & \shortstack{RSA\\(Ours)} \\

\midrule
\multicolumn{11}{@{}l}{\textbf{Llama-3-8B}} \\
Base & \underline{0.78} & 92.19 & 98.74 & 79.25 & 92.19 & \textbf{67.50} & \underline{64.75} & \underline{66.00} & \textbf{66.00} & 64.25 \\
LAT & \textbf{0.00} & 85.16 & 94.97 & 80.50 & 86.72 & \underline{66.50} & \underline{64.75} & 63.75 & 64.25 & 62.75 \\
CAT & \textbf{0.00} & 27.34 & \underline{49.69} & 80.50 & 45.31 & 65.75 & \textbf{65.75} & 61.00 & \underline{64.75} & 64.00 \\
MixAT & \textbf{0.00} & 78.91 & 83.02 & 87.42 & 82.03 & 62.25 & 63.50 & 57.25 & 62.75 & 60.25 \\
ReFAT & \textbf{0.00} & 89.84 & 97.48 & 72.96 & 92.19 & 66.25 & \textbf{65.75} & 64.25 & \underline{64.75} & \textbf{65.25} \\
DeepRefusal & \textbf{0.00} & \textbf{0.00} & \textbf{0.00} & \underline{22.64} & \underline{23.44} & 65.50 & \underline{64.75} & \textbf{66.75} & 63.00 & 64.25 \\

\textbf{Ours(ReSAT)} & \textbf{0.00} & \underline{4.00} & \textbf{0.00} & \textbf{0.00} & \textbf{11.20} & 65.75 & 63.75 & 65.50 & \textbf{66.00} & \underline{64.75} \\

\midrule
\multicolumn{11}{@{}l}{\textbf{Mistral-7B}} \\
Base & 29.69 & 69.53 & 29.69 & 88.68 & 69.53& \textbf{60.50} & \textbf{59.50} & 60.50 & \underline{52.00} & \textbf{59.50} \\
LAT & 3.91 & 62.50 & \underline{18.24} & 81.13 & 81.25 & 54.75 & 55.25 & 45.00 & 32.00 & 54.00 \\
CAT & 1.56 & 10.94 & 76.73 & 63.52 & 75.78 & 54.50 & 52.50 & 44.75 & 29.75 & 53.25 \\
MixAT & \underline{0.78} & \underline{6.25} & \textbf{0.00} & \textbf{4.40} & 65.62 & 54.50 & 54.75 & 47.50 & 47.25 & 55.75 \\
ReFAT & \textbf{0.00} & 28.91 & 87.42 & 49.69 & \underline{28.91} & \underline{59.00} & 57.25 & 53.00 & 50.50 & 57.25 \\
DeepRefusal & 3.12 & 78.91 & 50.94 & 89.94 & 78.91 & \underline{59.00} & \underline{57.75} & \textbf{55.75} & \textbf{52.75} & \underline{57.75} \\



\textbf{Ours(ReSAT)} & \textbf{0.00} & \textbf{1.60} & \textbf{0.00} & \underline{28.57} & \textbf{20.00} & 57.25 & 57.25 & \underline{55.50} & 47.25 & 57.25 \\

\midrule
\multicolumn{11}{@{}l}{\textbf{Qwen2.5-14B}} \\ Base & 1.56 & 79.69 & 96.23 & 98.11 & 79.69 & 76.50 & 76.50 & \textbf{78.50} & \underline{76.75} & \underline{76.50} \\ LAT & \textbf{0.00} & 47.66 & 83.65 & 95.60 & 47.66 & 76.75 & 76.00 & 76.25 & 76.00 & 76.00 \\ CAT & \underline{0.78} & 1.56 & 58.49 & \underline{3.14} & \textbf{1.56} & 74.75 & 75.50 & 75.50 & 73.00 & 75.50 \\ MixAT & \textbf{0.00} & \textbf{0.00} & \textbf{1.26} & \textbf{0.00} & 29.69 & 76.50 & 76.50 & \underline{77.25} & \textbf{78.00} & 70.25 \\ ReFAT & \textbf{0.00} & 27.34 & 98.11 & 96.86 & 64.06 & 76.75 & 76.50 & \textbf{78.50} & 76.50 & 75.50 \\ DeepRefusal & \textbf{0.00} & 61.72 & 96.23 & 86.16 & 72.66 & 76.25 & 76.25 & 75.75 & 76.50 & 75.75 \\





\textbf{Ours(ReSAT)} & \textbf{0.00} & \underline{1.50} & \underline{7.14} & 22.73 & \underline{8.80} & \underline{78.50} & \textbf{79.00} & \textbf{78.50} & \underline{76.75} & \textbf{79.00} \\

\bottomrule
\end{tabular}
\end{adjustbox}
\end{table*}
\endgroup

\begingroup
\begin{table*}[t]\caption{Jailbreak robustness and utility of base models and their safety-trained variants (\%).}
\label{tab:main-jailbreak-utility-results}
\centering
\normalsize
\setlength{\tabcolsep}{2pt}
\renewcommand{\arraystretch}{1}
\begin{adjustbox}{max width=\linewidth}
\begin{tabular}{@{}l*{8}{c}!{\hspace{2pt}\vrule width 0.4pt\hspace{2pt}}*{3}{c}@{}}
\toprule
& \multicolumn{8}{c}{Jailbreak safety score $\downarrow$} & \multicolumn{3}{c}{\shortstack{Utility}} \\
\cmidrule(lr){2-9}\cmidrule(lr){10-12}
Model & NoAtk & \shortstack{Auto\\DAN} & \shortstack{Human\\JB} & GCG & PAIR & Inp. & Soft & RFA & \shortstack{MMLU\\$\uparrow$} & \shortstack{GSM\\8K$\uparrow$} & \shortstack{XS\\Test$\uparrow$} \\
\midrule
\multicolumn{12}{@{}l}{\textbf{Llama-3-8B}} \\
Base & 1.57 & 18.20 & 2.20 & 22.97 & 28.70 & 86.61 & 79.20 & 62.31 & 66.43 & \textbf{76.19} & \textbf{99.00} \\
LAT & 0.04 & 1.74 & \underline{0.03} & 7.32 & 23.30 & 62.92 & \textbf{75.55} & 70.35 & \textbf{66.55} & 66.26 & 67.00 \\
CAT & \textbf{0.02} & \textbf{0.02} & \textbf{0.02} & \underline{6.57} & 32.51 & 59.07 & 80.26 & \textbf{33.12} & 65.08 & 70.28 & 26.00 \\
MixAT & \textbf{0.02} & \textbf{0.02} & \textbf{0.02} & 13.82 & \underline{1.95} & 66.79 & 77.34 & 60.97 & 63.70 & 71.80 & 49.00 \\
ReFAT & 0.44 & 21.43 & 1.16 & 28.72 & 28.29 & 83.00 & 82.67 & \underline{49.06} & \underline{66.54} & \underline{74.07} & \underline{91.00} \\
DeepRefusal & \underline{0.03} & \textbf{0.02} & \textbf{0.02} & 10.65 & 6.01 & \underline{52.26} & \underline{77.26} & 57.75 & 65.99 & 71.87 & 48.00 \\
\textbf{Ours(ReSAT)} & 0.04 & \underline{0.04} & 0.04 & \textbf{1.05} & \textbf{1.19} & \textbf{47.31} & 82.12 & 57.25 & 66.49 & 73.09 & 53.00 \\
\midrule
\multicolumn{12}{@{}l}{\textbf{Mistral-7B}} \\
Base & 37.79 & 79.02 & 65.56 & 79.96 & 62.60 & 94.51 & 83.28 & 78.31 & \underline{58.40} & \underline{41.85} & \underline{95.00} \\
LAT & 0.60 & 47.53 & 10.51 & 28.82 & 61.65 & 91.18 & 81.51 & 79.69 & 52.79 & 0.00 & 67.00 \\
CAT & 0.40 & \underline{2.32} & 0.33 & \underline{20.99} & 47.33 & 86.27 & \textbf{80.43} & 76.94 & 51.53 & 33.97 & 86.00 \\
MixAT & \underline{0.05} & \textbf{0.02} & \textbf{0.02} & \textbf{12.50} & \textbf{13.14} & \underline{77.46} & 80.78 & \underline{72.35} & 52.93 & 34.27 & 62.00 \\
ReFAT & 0.07 & 54.00 & 6.53 & 67.40 & 38.67 & 87.76 & 82.77 & 75.77 & 58.09 & 27.29 & 72.00 \\
DeepRefusal & 1.77 & 68.82 & 14.88 & 69.39 & 59.67 & 90.88 & 82.93 & 78.24 & 58.13 & 41.32 & \textbf{97.00} \\
\textbf{Ours(ReSAT)} & \textbf{0.04} & 4.12 & \underline{0.07} & 53.83 & \underline{30.49} & \textbf{64.63} & \underline{80.51} & \textbf{47.29} & \textbf{59.22} & \textbf{42.99} & 57.00 \\
\midrule
\multicolumn{12}{@{}l}{\textbf{Qwen2.5-14B}} \\ Base & 3.04 & 29.72 & 11.29 & 12.89 & 31.01 & 87.54 & 84.46 & 29.28 & 78.64 & 78.85 & \textbf{99.00} \\ LAT & 0.08 & 0.04 & \underline{0.06} & \textbf{0.26} & \underline{21.50} & 70.24 & 74.20 & \textbf{25.01} & 78.69 & 56.25 & 69.00 \\ CAT & 0.16 & \textbf{0.02} & \textbf{0.02} & \underline{1.62} & 35.81 & \underline{64.68} & \underline{62.25} & 39.89 & 77.45 & 82.94 & 44.00 \\ MixAT & \textbf{0.02} & \textbf{0.02} & \textbf{0.02} & 3.36 & \textbf{5.50} & \textbf{56.64} & \textbf{54.60} & 40.33 & 78.23 & 83.09 & 41.00 \\ ReFAT & 0.16 & 2.08 & 1.60 & 3.45 & 27.34 & 83.78 & 85.29 & 31.67 & \textbf{78.95} & 79.91 & 80.00 \\ DeepRefusal & 0.12 & \underline{0.03} & 0.08 & 32.58 & 43.18 & 77.74 & 84.81 & \underline{27.83} & 78.35 & 82.79 & \underline{90.00} \\
\textbf{Ours(ReSAT)} & \underline{0.04} & 4.58 & 0.65 & 52.48 & 44.21 & 80.60 & 83.75 & 52.07 & \underline{78.93} & \underline{85.90} & 47.00 \\
\bottomrule
\end{tabular}
\end{adjustbox}
\end{table*}
\endgroup

\begin{table*}[!t]
\caption{Llama-3-8B ablations (\%). RSA reports HarmBench ASR. Setup and mixture details are in Appendix~\ref{app:ablation-study-details}.}
\label{tab:training-recipe-ablation}
\label{tab:ah-config-summary}
\label{tab:ah-key-results}
\centering
\normalsize
\setlength{\tabcolsep}{2pt}
\renewcommand{\arraystretch}{1}
\begin{adjustbox}{max width=\linewidth}
\begin{tabular}{@{}l*{6}{c}@{}}
\toprule
Setup & MMLU$\uparrow$ & XSTest$\uparrow$ & AutoDAN$\downarrow$ & GCG$\downarrow$ & Inpainting$\downarrow$ & \shortstack{RSA$\downarrow$} \\
\midrule
\multicolumn{7}{@{}l}{\textbf{(i) SFT initialization: same SafeTrain mixture}} \\
Original SFT & 65.88 & 9.00 & \underline{1.76} & \underline{2.78} & \underline{49.93} & \textbf{7.20} \\
Mixed SFT & \underline{66.42} & \underline{46.00} & \textbf{0.04} & \textbf{2.35} & \textbf{46.38} & \underline{12.00} \\
No initial SFT & \textbf{66.56} & \textbf{97.00} & 12.71 & 18.91 & 84.41 & 93.60 \\
\midrule
\multicolumn{7}{@{}l}{\textbf{(ii) Input attacks: matched hard-benign exposure and updates}} \\
No input attacks & \textbf{66.53} & \textbf{65.00} & 9.94 & 7.86 & 67.77 & 16.80 \\
With input attacks & 66.49 & 53.00 & \textbf{0.04} & \textbf{1.05} & \textbf{47.31} & \textbf{11.20} \\
\midrule
\multicolumn{7}{@{}l}{\textbf{(iii) Hard-benign supervision: ordinary-utility control}} \\
No Hard Benign & 65.61 & 9.00 & \textbf{0.04} & 5.81 & \textbf{45.63} & \textbf{4.00} \\
With Hard Benign & \textbf{66.49} & \textbf{53.00} & \textbf{0.04} & \textbf{1.05} & 47.31 & 11.20 \\
\midrule
\multicolumn{7}{@{}l}{\textbf{(iv) Training rounds: successive checkpoints of the same run}} \\
Ours after round 1 & 66.38 & \textbf{56.00} & \textbf{0.04} & 3.50 & \underline{42.63} & 35.20 \\
Ours after round 2 & \textbf{66.54} & 45.00 & \textbf{0.04} & \textbf{0.04} & \textbf{40.34} & \underline{12.00} \\
Ours after round 3 & \underline{66.49} & \underline{53.00} & \textbf{0.04} & \underline{1.05} & 47.31 & \textbf{11.20} \\
\midrule
\multicolumn{7}{@{}l}{\textbf{(v) Complementary directions: fixed versus cumulative ablation}} \\
Repeated $d_1$ & 66.46 & \textbf{60.00} & \textbf{0.04} & 5.01 & 48.17 & 31.20 \\
Cumulative directions (Ours) & \textbf{66.49} & 53.00 & \textbf{0.04} & \textbf{1.05} & \textbf{47.31} & \textbf{11.20} \\
\midrule
\multicolumn{7}{@{}l}{\textbf{(vi) Training-time RSA: matched initialization, quotas, and updates}} \\
Without training RSA & \textbf{66.49} & \textbf{79.00} & \textbf{0.04} & 3.37 & 53.60 & 61.60 \\
With training RSA & \textbf{66.49} & 53.00 & \textbf{0.04} & \textbf{1.05} & \textbf{47.31} & \textbf{11.20} \\
\bottomrule
\end{tabular}
\end{adjustbox}
\end{table*}

\subsection{Experimental Setup}
\label{sec:experiment-setup} \paragraph{Models and evaluation sources.} The three-backbone comparison uses Llama-3-8B-Instruct \citep{dubey2024llama3}, Mistral-7B-Instruct \citep{jiang2023mistral}, Qwen2.5-14B-Instruct \citep{qwen2024qwen25}
Evaluation cohorts are documented in Appendix~\ref{app:main-table-protocol}.  We additionally rerun ReSAT with three independent seeds for each of the three model families; the results are consistent with the main-table findings, with full per-seed results reported in Appendix~\ref{app:seed-reproducibility}. We also conducted additional experiments with Gemma-7B-it \citep{gemmateam2024gemma}, with results reported in Appendix~\ref{app:gemma-additional-results}. 

\textbf{Refusal-Subspace Ablation evaluation.}
After training, we freeze the checkpoint and construct evaluation subspaces on the safety-trained model. The first direction is selected from the safety-trained model checkpoint or transferred from the unmodified instruction-tuned base model. We then add complementary directions found on the evaluated checkpoint and jointly remove their accumulated span, expanding from one to at most four directions under the declared constraints. Search stops when no eligible direction remains under KL constraints.

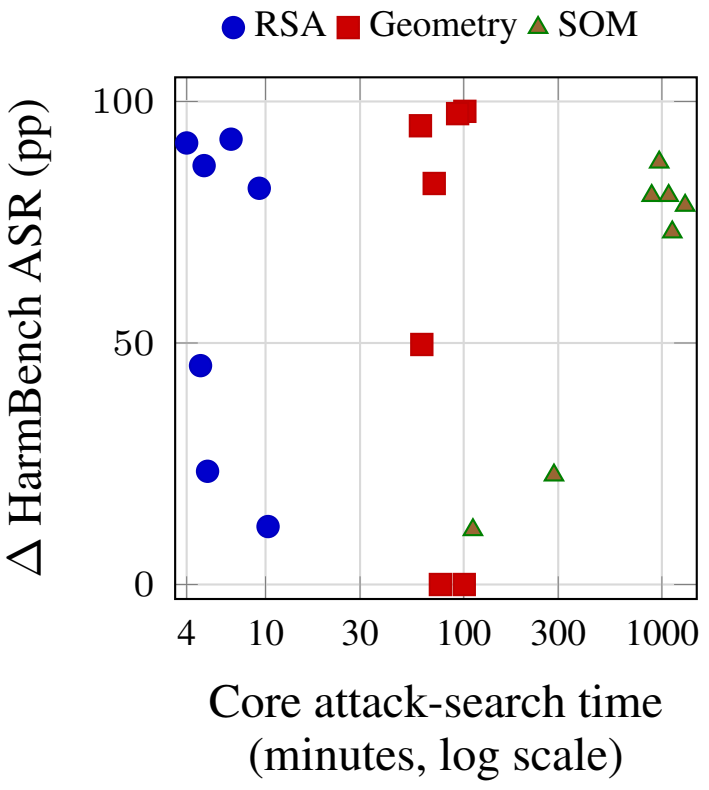
\begin{figure}[!b]
\centering
\begin{tikzpicture}
\begin{axis}[
  scale only axis,
  width=3.45cm,
  height=3.45cm,
  font=\footnotesize,
  xmode=log,
  log basis x=10,
  xmin=3.5,
  xmax=1500,
  ymin=-3,
  ymax=105,
  xlabel={Core attack-search time\\(minutes, log scale)},
  xlabel style={align=center},
  ylabel={$\Delta$ HarmBench ASR (pp)},
  xtick={4,10,30,100,300,1000},
  xticklabels={4,10,30,100,300,1000},
  tick label style={font=\scriptsize},
  grid=both,
  minor grid style={gray!15},
  major grid style={gray!30},
  legend columns=3,
  legend style={at={(0.5,1.02)},anchor=south,draw=none,font=\scriptsize},
  mark size=2pt,
]
\addplot+[only marks,color=blue!75!black,mark=*] coordinates {
  (4.0,91.41) (4.9,86.72) (4.7,45.31) (9.3,82.03)
  (6.7,92.19) (5.1,23.44) (10.3,12.00)
};
\addplot+[only marks,color=red!75!black,mark=square*] coordinates {
  (101.35,97.96) (60.50,94.97) (61.48,49.69) (71.02,83.02)
  (93.20,97.48) (100.67,0.00) (76.30,0.00)
};
\addplot+[only marks,color=green!50!black,mark=triangle*] coordinates {
  (1310.37,78.47) (1082.12,80.50) (888.52,80.50) (969.40,87.42)
  (1128.52,72.96) (285.48,22.64) (111.33,11.32)
};
\legend{RSA,Geometry,SOM}
\end{axis}
\end{tikzpicture}
\setlength{\abovecaptionskip}{3pt}
\caption{\footnotesize Core search time versus HarmBench ASR increase on seven Llama checkpoints; details are in Appendix~\ref{app:cost-multidirection}.}
\label{fig:cost-effectiveness}
\end{figure}

\textbf{Jailbreak attacks.} NoAtk denotes the unmodified harmful-prompt control, not a separate attack method. We evaluate AutoDAN genetic prompt search \citep{liu2024autodan}, GCG \citep{zou2023gcg}, and PAIR \citep{chao2023pair}. HumanJB uses the released HarmBench human-written jailbreak templates \citep{mazeika2024harmbench}, sampled five per behavior in our evaluation. Inp. denotes a fixed transfer corpus of diffusion-inpainting attacks \citep{ludke2025inpainting}: we use the third-party Circuit Breakers attack-completions release \citep{samuelsimko2026circuitrr}, not a new run of the original paper's complete search procedure. Soft uses the enhanced continuous soft-prompt implementation \citep{schwinn2024revisiting}. RFA uses the layerwise, mean-centered refusal-feature ablation operator evaluated by ReFAT \citep{yu2024refat}; it is distinct from our adaptive joint-subspace attack.


\textbf{Computational cost.} Single-run wall-clock times on NVIDIA GH200 120GB GPUs are 154.2 minutes for SFT, 387.6 for SafeTrain, and 619.2 end to end for Llama-3-8B ReSAT. Training uses two GPUs, the second for PAP generation. Figure~\ref{fig:cost-effectiveness} compares seven Llama checkpoints; RSA has the lowest measured core-search time, and on Base RSA/Geometry/SOM take 4/101/1,310 minutes. Its low cost enables repeated SafeTrain direction refreshes, whereas the other joint-ablation searches are substantially costlier. Full accounting and attack-specific timing boundaries are in Appendix~\ref{app:computational-cost}.

\textbf{Scoring and utility.} Attack Successful Rate (ASR) is measureed by calculating average of 3 classifier: HarmBench \citep{mazeika2024harmbench}, Llama Guard 2 \citep{metallamaguard2}, and StrongREJECT \citep{souly2024strongreject}, reported as percentages. MMLU \citep{hendrycks2021mmlu} and GSM8K \citep{cobbe2021gsm8k} measure capability; XSTest \citep{rottger2024xstest} measures benign compliance. Clean has no direction intervention; lexical RR is Refusal Rate reported as another supplementary metrics in Appendix~\ref{app:lexical-refusal-rates}. Best values are \textbf{bold}; second-best are \underline{underlined}. 

\par



\subsection{Resistance to Refusal Ablation Attacks}
\label{sec:resistance-jailbreak}
\label{sec:brittleness}
\label{sec:deeprefusal-refat-comparison}
\textbf{ReSAT provides strong defense against a range of refusal-ablation attacks.} Table~\ref{tab:main-three-family-results} shows strong resistance to both single-direction and different types of cumulative joint ablation across 3 model backbones, with substantially lower ASR than several existing safety-training baselines.

\textbf{Although ReSAT uses RSA to optimize SafeTrain, it also provides strong defense against Geometry and SOM.} For example, Llama Ours achieves 0.00\% ASR under both attacks (Table~\ref{tab:main-three-family-results}), showing that robustness learned with RSA can be transferred to other attacks. This transfer demonstrates good generalization and robustness across different models 

\subsection{Attack Performance}
\label{sec:attack-analysis}
\textbf{RSA, our joint-ablation attack, is highly effective against existing safety-trained models.} Models that resist single-direction ablation can remain vulnerable to joint removal of multiple directions. For example, Mistral MixAT's ASR rises from 6.25\% under Single to 65.62\% under rank-three RSA; similar weaknesses appear in CAT, DeepRefusal and other safety-training baselines (Table~\ref{tab:main-three-family-results}). Relaxing the KL constraint can strengthen RSA but may incur greater utility loss, with effects varying across models (Appendix~\ref{app:kl-ablation}).

\textbf{RSA is more time-efficient than Geometry and SOM in our measured runs (Figure~\ref{fig:cost-effectiveness}).} In the Llama timing study, our core search takes 10.3 minutes, versus 76.30 for Geometry and 111.33 for SOM. Although search budgets differ, RSA's low measured cost makes the attack practical and enables repeated direction refreshes during SafeTrain. Together, these results show that single-direction robustness is not a reliable proxy for subspace robustness, while RSA exposes this gap at substantially lower measured search cost.

\subsection{Utility Preservation and Jailbreak Robustness}
\label{sec:utility-jailbreak}
\textbf{ReSAT largely preserves general capability after SafeTrain.} In Table~\ref{tab:main-jailbreak-utility-results}, MMLU for Llama Ours (66.49\%) and Base (66.43\%) ; for Mistral Ours(59.22\%)/Base(58.40\%); Qwen ReSAT(ours) achieves 78.93\% versus base 78.64\% with also GSM8K improves from 78.85\% to 85.90\%  relativr to base. ReSAT exhibits over-refusal on XSTest, a limitation also observed in other baseline CAT, MixAT, and DeepRefusal. However, On Llama-3-8B, Our ReSAT achieves higher benign compliance.

\textbf{ReSAT also defends effectively against several jailbreak attacks, outperforming ReFAT and DeepRefusal in multiple tested settings.} On Llama, Ours's GCG/PAIR/Inpainting mixed safety scores (lower is better) are 1.05/1.19/47.31, compared with 28.72/28.29/83.00 for ReFAT and 10.65/6.01/52.26 for DeepRefusal. On Mistral, Ours's AutoDAN score is 4.12, versus 54.00 for ReFAT and 68.82 for DeepRefusal (Table~\ref{tab:main-jailbreak-utility-results}). These gains vary across attacks, but they show promising robustness for our safe-training method ReSAT.

\subsection{Ablation Study}
\label{sec:ablation-study}
\label{sec:ablation-recipes}
\label{sec:ablation-recipe-results}


\textbf{SFT initialization affects the robustness compliance trade-off.} Table~\ref{tab:training-recipe-ablation}(i) compares original, mixed, and no SFT initialization under the same three-round SafeTrain mixture; total training compute differs. Mixed SFT improves benign compliance but increases observed RSA ASR, while omitting SFT largely removes RSA resistance (Appendix~\ref{app:ablation-study-details}).

\textbf{Input-attack training improves measured jailbreak robustness.} Group (ii) removes continuous/PAP attacks from both SFT and SafeTrain while preserving hard-benign exposure and update budgets. Removing these attacks raises the Inpainting score from 47.31\% to 67.77\%, testing their combined contribution across both stages.

\textbf{Hard-benign supervision improves benign compliance.} Group (iii) replaces hard-benign examples with ordinary utility while holding the harmful branches and update budget fixed. Adding hard-benign supervision raises XSTest compliance from 9\% to 53\%, at the cost of higher rank-four RSA ASR.

\textbf{RSA resistance improves mainly between the first two rounds.} Group (iv) compares checkpoints after rounds one, two, and three of the sametraining run. Rank-four ASR falls from 35.20\% to 12.00\% and 11.20\%, with only a one-response difference between the final two rounds (Appendix~\ref{app:ablation-study-details}).

\textbf{Cumulative-direction training improves resistance over repeated single-direction training.} Group (v) compares repeated ablation of the first fixed direction $d_1$ with cumulative direction ablation during training, using the same final-model evaluation protocol. Cumulative training lowers rank-four ASR from 31.20\% to 11.20\%.

\textbf{Training-time RSA improves refusal-ablation resistance in a matched control.} Group (vi) disables only training-time direction ablation while retaining the initialization, sample quotas, continuous/PAP attacks, and update budget; both models are evaluated with RSA. At the shared attack rank three, training-time RSA lowers ASR from 61.60\% to 12.00\%, with lower benign compliance.

\subsection{Mechanism Analysis}\label{sec:mechanism-analysis}

\begin{figure}[tbp]
  \centering
  \includegraphics[
    width=\linewidth,
    trim=2mm 1mm 2mm 1mm,
    clip
  ]{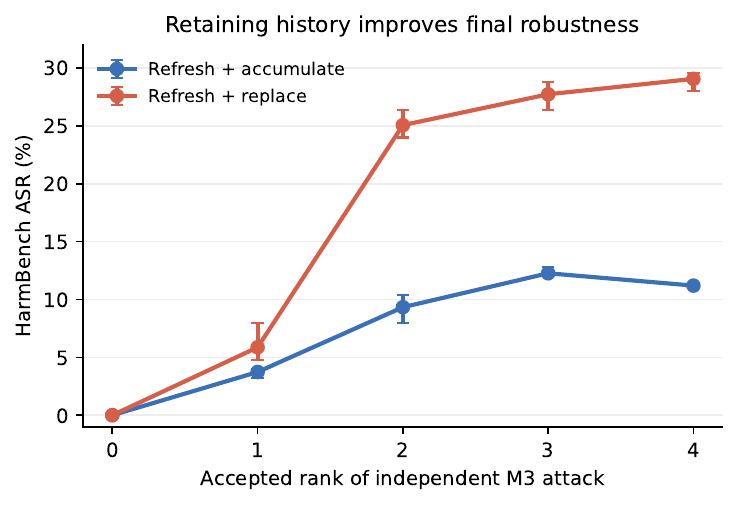}
  \caption{History retention lowers HarmBench ASR across accepted ranks of an independent $M_3$ attack compared with replacing the previous direction bank (\%; lower is better): History retention improves robustness}
  \label{fig:history-retention}
\end{figure}

    \textbf{SafeTrain progressively suppresses attacks from known direction banks.} Appendix Figure~\ref{fig:cumulative-mechanism}A shows that the causal attack effect of each identified bank decreases across successive checkpoints. For example, the ASR increase induced by $B_1$ falls from 51.2 pp on $M_0$ to 8.0, 1.9, and 2.4 pp on $M_1$, $M_2$, and $M_3$, demonstrating robustness acquisition against known interventions.

\textbf{Retaining direction history improves final robustness.} We compare refresh+accumulate with a matched refresh+replace schedule that re-estimates a direction every round but discards the preceding bank. Replacement reaches 29.07\% HarmBench ASR on $M_3$, compared with 11.20\% under accumulation (Figure~\ref{fig:history-retention}), a paired difference of 17.87 pp (95\% CI: $[14.13,21.87]$).



\textbf{Cumulative history prevents rediscovery of covered directions.} After projection against the historical span, accumulated $d_2$ and $d_3$ retain 76.0\% and 99.9\% of their squared energy, whereas replacement candidates retain only 0.185\% and 0.043\% (Appendix Figure~\ref{fig:cumulative-mechanism}C). Thus, accumulation exposes complementary information while replacement largely rediscovers covered directions.

\textbf{The complementary directions are causally active.} Adding accumulated $d_2$ and $d_3$ to their preceding banks increases held-out ASR by 26.93 and 4.27 pp, respectively, while replacement candidates provide no positive marginal gain (Appendix Figure~\ref{fig:cumulative-mechanism}D). The newly exposed information therefore causally controls refusal rather than merely reflecting representation rotation.

\section{Conclusion} \label{sec:conclusion} Resistance to a single refusal direction does not ensure safety under joint ablation. RSA adaptively discovers and jointly removes complementary directions, achieving competitive attack effectiveness with substantially lower measured search cost than Geometry and SOM. ReSAT combines mixed SFT with safety training under cumulative ablation. Across model families, ReSAT lowers attack success rates relative to evaluated baselines in key settings while preserving general capability. Its robustness extends to other refusal-ablation and several jailbreak attacks, complementing existing adversarial training for open-weight models. Our mechanism analysis further shows that retaining the cumulative direction history prevents rediscovery of already-covered features and exposes causally active complementary directions.

\par \section*{AI use statement}
We used generative AI tools to assist with translation, language editing, drafting and revising portions of the manuscript, interpreting experimental results, and checking consistency between the text and tables. These tools also assisted with LaTeX editing and formatting. 

\section*{Ethics statement}
This work studies refusal-ablation attacks to evaluate and improve the safety of open-weight language models. RSA has dual-use implications: interventions that reveal vulnerabilities could also be misused to bypass safeguards and elicit harmful content. Our intended use is controlled safety evaluation and the development of defenses such as ReSAT. The reported robustness gains are limited to the evaluated models, attacks, and benchmarks and do not constitute a general safety guarantee. We encourage responsible use of these methods for authorized safety research and defensive evaluation.

\bibliography{references,citation_additions_20260914}

\clearpage
\appendix
\setlength{\intextsep}{8pt plus 1pt minus 1pt}
\onecolumn
\raggedbottom
\begin{multicols}{2}
\raggedcolumns

\section{Detailed Refusal Ablation Protocol}
\label{app:attack-details}

\subsection{Background: Naive Refusal Ablation}

Let $h_{p,\ell}^{M}(x)\in\mathbb{R}^{d}$ denote the clean residual activation at the input of decoder layer $\ell$ for prompt $x$ in model $M$. Here $d$ is the hidden dimension, $p\in\mathcal P$ is a candidate prompt position relative to the assistant boundary, and $\ell\in\{0,\ldots,L-1\}$ is a zero-based layer index in a model with $L$ decoder layers. Given harmful and harmless extraction sets $\mathcal D_{\mathrm{harm}}^{\mathrm{ext}}$ and $\mathcal D_{\mathrm{safe}}^{\mathrm{ext}}$, we first compute clean difference-in-means candidates:
\begin{equation}
\begin{aligned}
c_{p,\ell}^{M}
&=
\frac{1}{|\mathcal D_{\mathrm{harm}}^{\mathrm{ext}}|}
\sum_{x\in\mathcal D_{\mathrm{harm}}^{\mathrm{ext}}}h_{p,\ell}^{M}(x)\\
&\quad-
\frac{1}{|\mathcal D_{\mathrm{safe}}^{\mathrm{ext}}|}
\sum_{x\in\mathcal D_{\mathrm{safe}}^{\mathrm{ext}}}h_{p,\ell}^{M}(x).
\end{aligned}
\label{eq:mean-difference}
\end{equation}
For a nonzero candidate, let $d_{p,\ell}=c_{p,\ell}^{M}/\|c_{p,\ell}^{M}\|_2$. For any unit direction $d$, rank-one ablation removes its component from an intervened activation $h$:
\begin{equation}
A_d(h)=h-(h^\top d)d.
\label{eq:rank1-ablation}
\end{equation}
The source pair $(p,\ell)$ specifies where the candidate is extracted, not where it is removed. Unless stated otherwise, the same projection is applied at all sequence positions and decoder layers, at the residual input and the attention and MLP outputs, throughout prompt processing and generation. A rank-one attack keeps this one direction fixed; it is not a one-token or one-forward-pass intervention.

\paragraph{First-output-token selection.}
Candidate extraction and candidate selection are separate operations. Our naive selector fixes
\begin{equation}
R=\{t_I\},
\label{eq:naive-probe}
\end{equation}
where $t_I$ is the token ID corresponding to \texttt{I} at the first assistant-output position under the model's tokenizer and chat template. The probe is evaluated immediately after the native assistant boundary, without inserting \texttt{I} into the response. In particular, a refusal direction is not the embedding of \texttt{I}.

For a general set $R$ of distinct probe token IDs, let $p_{M,Q}(\cdot\mid x)$ denote the distribution of the first token to be generated when the subspace with orthonormal basis $Q$ is ablated; $Q=\varnothing$ denotes the clean model. Using harmful and harmless validation sets $\mathcal D_{\mathrm{harm}}^{\mathrm{val}}$ and $\mathcal D_{\mathrm{safe}}^{\mathrm{val}}$, we define
\begingroup\small
\begin{align}
m_R(x;M,Q)
&=\log\frac{\sum_{t\in R}p_{M,Q}(y_1=t\mid x)}
{1-\sum_{t\in R}p_{M,Q}(y_1=t\mid x)},
\label{eq:first-token-margin}\\
S_R(Q;M)
&=\frac{1}{|\mathcal D_{\mathrm{harm}}^{\mathrm{val}}|}
\sum_{x\in\mathcal D_{\mathrm{harm}}^{\mathrm{val}}}m_R(x;M,Q),
\label{eq:bypass-score}\\
K(Q;M)
&=\frac{1}{|\mathcal D_{\mathrm{safe}}^{\mathrm{val}}|}
\sum_{x\in\mathcal D_{\mathrm{safe}}^{\mathrm{val}}}\notag
\\[-0.3ex]
&\;\KL\!\left(p_{M,\varnothing}(\cdot\mid x)\,\|\,
p_{M,Q}(\cdot\mid x)\right).
\label{eq:harmless-kl}
\end{align}
\endgroup
Thus $m_R$ takes the log-odds of the \emph{sum} of probe-token probabilities, and $S_R$ averages these per-prompt log-odds. The KL is over the full vocabulary at the first output step, from the clean distribution to the distribution under the complete intervention, not between consecutive attack rounds. $S_R$ is a direction-selection proxy, not the refusal rate (RR) measured from generated answers.

Let $\mathcal E_1(R;b_{\mathrm{add}})$ be the eligible source pairs defined below for an empty initial basis, where $b_{\mathrm{add}}\in\{0,1\}$ records whether the optional activation-addition check is enabled. The naive attack selects
\begin{equation}
(p^\star,\ell^\star)\in
\arg\min_{(p,\ell)\in\mathcal E_1(\{t_I\};b_{\mathrm{add}})}
S_{\{t_I\}}([d_{p,\ell}];M),
\label{eq:naive-selection}
\end{equation}
and retains that fixed rank-one projection throughout generation. The standalone \texttt{I}-selection implementation uses the KL and source-layer constraints without the addition check ($b_{\mathrm{add}}=0$). Addition-constrained runs are specified separately. The \texttt{Sorry} and \texttt{I+Sorry} selectors replace $R$ by $\{t_{\mathrm{Sorry}}\}$ and $\{t_I,t_{\mathrm{Sorry}}\}$, respectively; they are probe variants, not different definitions of candidate extraction.

\subsection{Refusal Subspace Ablation}

Refusal Subspace Ablation adaptively selects complementary directions and jointly removes their span while keeping model weights fixed. Let $B_0\in\mathbb{R}^{d\times0}$ be empty. At step $k$, $B_{k-1}$ has orthonormal columns spanning the components already removed. The clean candidate pool is
\begin{equation}
\mathcal C^M=\{c_{p,\ell}^{M}:p\in\mathcal P,\ \ell=0,\ldots,L-1\}.
\label{eq:adaptive-candidate}
\end{equation}
It is computed once from Equation~\ref{eq:mean-difference} and can be cached for this checkpoint. We do not recompute harmful/harmless activation means under the current joint intervention. Instead, every step re-residualizes and jointly re-scores the clean candidates against the current span:
\begin{equation}
\begin{aligned}
r_{k,p,\ell}&=(I_d-B_{k-1}B_{k-1}^{\top})c_{p,\ell}^{M},\\
\nu_{k,p,\ell}&=\frac{r_{k,p,\ell}}{\|r_{k,p,\ell}\|_2}.
\end{aligned}
\label{eq:adaptive-residual}
\end{equation}
Here $I_d$ is the $d\times d$ identity matrix; $r_{k,p,\ell}$ is the unnormalized residual, and $\nu_{k,p,\ell}$ is its unit vector, defined only for nonzero residuals. For each surviving candidate, form
\begin{equation}
Q_{k,p,\ell}=\operatorname{Basis}([B_{k-1},\nu_{k,p,\ell}]),
\label{eq:expanded-attack-basis}
\end{equation}
where $\operatorname{Basis}$ returns an orthonormal basis for the columns' span. Both $S_R(Q_{k,p,\ell};M)$ and $K(Q_{k,p,\ell};M)$ evaluate the complete expanded intervention, not the new vector in isolation.

\paragraph{Optional activation-addition check.}
Some selection protocols additionally require a candidate to induce the refusal probe on harmless validation prompts. Let $p^{+}_{M;r,\ell}(\cdot\mid x)$ be the first-output-token distribution obtained by adding the \emph{unnormalized} vector $r$ to the residual input of source layer $\ell$ at all prompt positions. This is a separate clean-model forward pass with only the addition hook: the current subspace is \emph{not} simultaneously ablated. Define
\begin{equation}
\begin{aligned}
J_R(r,\ell;M)&=
\frac{1}{|\mathcal D_{\mathrm{safe}}^{\mathrm{val}}|}
\sum_{x\in\mathcal D_{\mathrm{safe}}^{\mathrm{val}}}
\\[-0.3ex]
&\quad\log\frac{\sum_{t\in R}p^{+}_{M;r,\ell}(y_1=t\mid x)}
{1-\sum_{t\in R}p^{+}_{M;r,\ell}(y_1=t\mid x)}.
\end{aligned}
\label{eq:addition-score}
\end{equation}
When enabled, the check uses $r=r_{k,p,\ell}$, not $\nu_{k,p,\ell}$, and requires $J_R\geq\eta$. It is a probe-based steering check rather than a generated-answer RR test.

\paragraph{Eligibility and joint selection.}
For nonzero raw candidates, let $\mathcal E_k(R;b_{\mathrm{add}})$ contain the source pairs satisfying all of
\begin{align}
& p\in\mathcal P,\qquad
0\leq\ell<\left\lfloor(1-\rho_{\mathrm{prune}})L\right\rfloor,
\label{eq:attack-layer-cutoff}\\
& \|r_{k,p,\ell}\|_2>0,\qquad
\frac{\|r_{k,p,\ell}\|_2}{\|c_{p,\ell}^{M}\|_2}\geq\epsilon,
\notag\\
& K(Q_{k,p,\ell};M)\leq\tau,
\label{eq:attack-eligibility}\\
& b_{\mathrm{add}}=0\quad\text{or}\quad
J_R(r_{k,p,\ell},\ell;M)\geq\eta.
\label{eq:optional-addition-gate}
\end{align}
Here $\rho_{\mathrm{prune}}$ is the fraction of final source layers excluded, $\epsilon$ is the minimum retained residual fraction, and $\tau$ is the harmless first-token KL budget. We use $\rho_{\mathrm{prune}}=0.2$ unless stated otherwise: for $L=32$, the implementation admits $\ell<25$, not $\ell<25.6$. These are source-layer restrictions, not restrictions on where ablation is applied. The switch $b_{\mathrm{add}}$ and threshold $\eta$ specify the optional check. At $k=1$, $r_{1,p,\ell}=c_{p,\ell}^{M}$ and $Q_{1,p,\ell}=[d_{p,\ell}]$, which defines the naive eligibility set above (with $\epsilon\leq1$).

The selected component and updated basis are
\begin{equation}
\begin{aligned}
(p_k,\ell_k)&\in
\arg\min_{(p,\ell)\in\mathcal E_k(R;b_{\mathrm{add}})}
S_R(Q_{k,p,\ell};M),\\
B_k&=\operatorname{Basis}([B_{k-1},\nu_{k,p_k,\ell_k}]).
\end{aligned}
\label{eq:deep-selection}
\end{equation}
Writing $d_k=\nu_{k,p_k,\ell_k}$, joint ablation then removes the entire accumulated span:
\begin{equation}
A_{B_k}(h)=h-B_kB_k^\top h.
\label{eq:joint-ablation}
\end{equation}
The procedure repeats until the requested rank is reached or no candidate is eligible. It imposes no requirement that the proxy score or generated RR improve over the previous step, so neither is guaranteed to decrease monotonically. When the selector, prompt preprocessing and splits, candidate grid, and eligibility constraints are identical, the first step from $B_0=\varnothing$ equals the corresponding naive attack; subsequent steps add orthogonal components and retain joint removal of all selected directions. The direction bank is only an implementation of this subspace.

\paragraph{Protocol scope.}
The general adaptive-matrix runner defaults to $b_{\mathrm{add}}=1$ and $\eta=0$. Some Mistral continuation runs instead set $\eta=-10^9$, effectively disabling the addition filter; the standalone naive extractor omits it. Consequently, the label \emph{strict} alone does not establish that two runs use identical filters: comparisons must retain their recorded selector, KL budget, and addition-filter setting. Training-time selection for Llama H uses the \texttt{I} probe, KL $\leq0.2$, and no addition check ($b_{\mathrm{add}}=0$); the default evaluation bound is KL $\leq0.1$. The historical F training selector also omits the addition check. Appendix~\ref{app:training-details} gives the search settings. Section~\ref{sec:comparison} evaluates refusal and utility under the selected interventions.

\end{multicols}
\par\noindent\begin{minipage}{\textwidth}
\subsection{RSA Pseudocode}
\label{app:rsa-pseudocode}
Algorithm~\ref{alg:rsa} summarizes the executable RSA loop specified in Appendix~\ref{app:attack-details}. 


\begin{algorithm}[H]
\caption{Refusal Subspace Ablation (RSA).}
\label{alg:rsa}
\small
\begin{algorithmic}[1]
\Require Frozen model $M$; extraction sets $\mathcal D^{\rm ext}_{\rm harm},\mathcal D^{\rm ext}_{\rm safe}$; validation sets $\mathcal D^{\rm val}_{\rm harm},\mathcal D^{\rm val}_{\rm safe}$; maximum rank $K$; probe-token set $R$; KL budget $\tau$
\Ensure Orthonormal refusal-subspace basis $B$
\State Extract the clean candidate pool $\mathcal C^M=\{c^M_{p,\ell}\}_{p,\ell}$ by difference in means
\State $B\gets\varnothing$
\For{$k=1,\ldots,K$}
    \State $\mathcal E\gets\varnothing$
    \For{$c\in\mathcal C^M$}
        \State $r\gets(I-BB^\top)c$
        \If{$\lVert r\rVert_2$ is negligible} \State \textbf{continue} \EndIf
        \State $Q\gets\operatorname{Basis}([B,r/\lVert r\rVert_2])$
        \If{$K(Q;M)>\tau$} \State \textbf{continue} \EndIf
        \State Add $(S_R(Q;M),Q)$ to $\mathcal E$ \Comment{apply the optional addition filter if enabled}
    \EndFor
    \If{$\mathcal E=\varnothing$} \State \textbf{break} \EndIf
    \State $B\gets Q^\star$, where $(\cdot,Q^\star)\in\arg\min_{(s,Q)\in\mathcal E}s$
\EndFor
\State \Return $B$; throughout generation jointly ablate $h\gets h-BB^\top h$
\end{algorithmic}
\end{algorithm}
\end{minipage}
\par\vspace{10pt}
\begin{multicols}{2}
\raggedcolumns

\section{Training Details and Diagnostics}
\label{app:training-variants}
\label{app:bcde-summary}
This appendix describes the core training protocol and two retained mechanistic diagnostics. Ablation setups and sampling shares are in Appendix~\ref{app:ablation-study-details}; shared evaluation definitions and cohorts are in Appendix~\ref{app:main-table-protocol}.

\subsection{Core SFT and joint-ablation training protocol}
\label{app:training-details}
\begin{algorithm}[H]
\caption{ReSAT-H training.}
\label{alg:resat-h}
\small
\begin{algorithmic}[1]
\Require Base model $M$; mixed-SFT data $\mathcal D_{\rm SFT}$; SafeTrain data $\mathcal D$; rounds $J=3$
\Ensure Last completed LoRA adapter and retained direction bank
\State $\theta^{(0)}\gets\operatorname{MixedSFT}(M,\mathcal D_{\rm SFT})$ with all projections disabled
\State $B_0\gets\varnothing$
\For{$j=1,\ldots,J$}
    \State Freeze $M_{\theta^{(j-1)}}$ and refresh its clean RSA candidate pool
    \State $d_j\gets\operatorname{RSASelectOne}(M_{\theta^{(j-1)}},B_{j-1})$
    \If{no eligible $d_j$ exists} \State \textbf{break} \EndIf
    \State $B_j\gets\operatorname{Basis}([B_{j-1},d_j])$
    \State Set H/R/CR/PR/U/HB sampling shares to $10/22.5/6.25/6.25/35/20\%$
    \State Apply $B_j$ in the R/CR/PR branches; keep U/HB branches unablated
    \State $\theta^{(j)}\gets\operatorname{SafeTrain}(\theta^{(j-1)},B_j,\mathcal D)$ for 32 updates
    \Statex \hspace{\algorithmicindent}\Comment{update LoRA only with response-only, sequence-normalized cross-entropy}
\EndFor
\State \Return the last completed adapter and direction bank
\end{algorithmic}
\end{algorithm}

Mixtures are presentation percentages in the order harmful/response-ablation/continuous-response-ablation/PAP-response-ablation/utility/hard-benign (H/R/CR/PR/U/HB).

Unless otherwise stated, this subsection describes the shared training mechanics used by the primary Llama configuration. It starts from the mixed-SFT checkpoint with hard-benign supervision and runs three SafeTrain rounds of 4,000 presentations (32 updates) using H/R/CR/PR/U/HB shares $10/22.5/6.25/6.25/35/20$. Before each round, RSA refreshes candidates on the current checkpoint and appends an eligible complementary direction. If a complete candidate sweep finds none, the original procedure stops and retains the last completed round; it does not automatically continue with the old bank. Llama ours completed all three rounds, with training-bank ranks $(1,2,3)$. Fixed-bank continuations are separate variants. Exact selector and eligibility definitions are in Appendix~\ref{app:attack-details}, while row-specific evaluation exceptions are in Appendix~\ref{app:main-table-protocol}.

For an orthonormal direction bank $B$, every intervened activation uses $A_B(h)=(I-BB^\top)h$. The pretrained weights are frozen and only LoRA parameters are updated. All conditions use response-only, sequence-normalized cross-entropy on the assistant target; harmful examples receive safe refusals, while utility and hard-benign examples retain helpful targets.

Mixed SFT uses clean harmful, continuous-attack, PAP-plus-continuous-attack, utility, and hard-benign branches with projection disabled. SafeTrain adds pure cumulative-subspace ablation and retains the current bank during both input-attack search and the outer safe-response update. Continuous perturbations are optimized only on user-content embeddings toward a separate unsafe prefix, aligned to the safe-target sequence, and detached before the outer update.

\end{multicols}
\par\noindent\begin{minipage}{\textwidth}
\begin{table}[H]
\caption{Forward conditions in SafeTrain round $j$. All harmful conditions supervise the same safe refusal. Continuous perturbations are aligned to the outer sequence and detached. In mixed SFT all projections are disabled and the pure-ablation condition is absent; SFT and SafeTrain sampling shares are given in Appendix~\ref{app:ablation-study-details}.}
\label{tab:f-training-conditions}
\centering
\normalsize
\setlength{\tabcolsep}{4pt}
\begin{adjustbox}{max width=\linewidth}
\begin{tabular}{lcccc}
\toprule
Condition & User input & Outer target & Projection & \shortstack{Input\\perturbation} \\
\midrule
Clean refusal & $x$ & $y_{\mathrm{ref}}$ & None & None \\
Pure ablation & $x$ & $y_{\mathrm{ref}}$ & $B_j$ & None \\
Continuous & $x$ & $y_{\mathrm{ref}}$ & $B_j$ & Continuous \\
PAP+continuous & $P(x;\xi)$ & $y_{\mathrm{ref}}$ & $B_j$ & Continuous \\
Utility & $x$ & $y_{\mathrm{util}}$ & None & None \\
Hard-benign & $x$ & $y_{\mathrm{ben}}$ & None & None \\
\bottomrule
\end{tabular}
\end{adjustbox}
\end{table}
\end{minipage}
\par
\begin{multicols}{2}
\raggedcolumns

\paragraph{Training-time direction selection.}
Llama ours uses the \texttt{I} first-output-token probe, a harmless first-token KL bound of $0.2$, and no activation-addition check. Each sweep uses 128 harmful and 128 harmless extraction prompts, with 32 validation prompts per pool (extraction seed 20260825; validation seed 20260826). Candidates cover the five assistant-boundary positions; the last 20\% of source layers are excluded, the minimum residual fraction is $10^{-3}$, and the basis tolerance is $10^{-5}$. Extraction uses the unquantized checkpoint. These training gates are distinct from the default evaluation KL bound of $0.1$.


\end{multicols}
\par\noindent\begin{minipage}{\textwidth}
\begin{table}[H]
\caption{ A SafeTrain round is one full pass over the constructed 4,000-presentation pool; it does not necessarily add a direction. All models first receive four mixed-SFT epochs (64 updates).}
\label{tab:f-training-schedules}
\centering
\normalsize
\begin{adjustbox}{max width=\linewidth}
\begin{tabular}{lccc}
\toprule
\shortstack[l]{Original instruction-tuned\\checkpoint} & \shortstack{SafeTrain\\rounds} & Updates & Ranks by round \\
\midrule
Llama-3-8B-Instruct & 3 & 96 & $1,2,3$ \\
Mistral-7B-Instruct-v0.2 & 2 & 64 & $1,2$ \\
Qwen2.5-14B-Instruct & 1 & 32 & $1,2$ \\
\bottomrule
\end{tabular}
\end{adjustbox}
\end{table}
\end{minipage}
\par
\begin{multicols}{2}
\raggedcolumns

\paragraph{Optimization settings for Llama ReSAT.}
Mixed SFT uses 2,000 presentations per epoch for four epochs (64 updates), with learning rate $2\times10^{-4}$; SafeTrain uses one 4,000-presentation epoch per round (32 updates), with learning rate $5\times10^{-6}$. Both stages use AdamW, microbatch size 2, nominal effective batch size 128, weight decay $10^{-3}$, gradient clipping at $0.3$, and two warm-up updates. LoRA has rank 64, scaling parameter 16, and dropout $0.1$, on the attention query/key/value/output and MLP gate/up/down projections, with no trainable bias. The frozen backbone uses NF4 with BF16 computation. The utility response cap is 256 tokens; harmful and hard-benign sequences are capped at 1,024 tokens. Native chat templates and response boundaries are validated.

The inner embedding attack takes 10 sign-gradient steps of size $10^{-4}$, with radius $0.075$ times the mean embedding norm, on user-content tokens only. PAP uses local Mixtral-8x7B-Instruct-v0.1 (NF4/BF16), at most 384 generated tokens, temperature $1.0$, top-$p=0.9$, and at most three attempts. Perturbations are detached for the outer response-only loss. The training seed is 20260908; harmful, utility, hard-benign, and evaluation pools are kept disjoint. 

\end{multicols}
\par\noindent\begin{minipage}{\textwidth}
\subsection{Mechanistic diagnostics retained from early experiments}
\label{app:early-training-diagnostics}

We retain the two diagnostics that directly test whether refusal wording and refusal-token choice change the learned direction. 

\subsubsection{Ten refusal patterns after SFT}
\label{app:multipattern}

A separate QLoRA SFT experiment pairs 100 harmful prompts with each of the ten targets in Table~\ref{tab:patterns} (1,000 refusal examples) and 1,000 UltraChat utility examples, with no direction ablation. Training uses four epochs, followed by two short calibration rounds that balance the output branches.

\begin{table}[H]
\caption{Ten refusal targets used for the multi-pattern SFT diagnostic.}
\label{tab:patterns}
\centering
\normalsize
\setlength{\tabcolsep}{3pt}
\begin{tabularx}{\linewidth}{>{\centering\arraybackslash}p{0.22in}X}
\toprule
ID & Target response \\
\midrule
1 & I cannot help with that request. I can instead provide safe, legal, and prevention-focused information. \\
2 & Sorry, but I am unable to provide those instructions. I can offer a high-level discussion of the risks or safer alternatives. \\
3 & Your request is denied because fulfilling it could enable harmful activity. Assistance with prevention or protection remains available. \\
4 & Providing the requested material is not allowed. Related defensive, educational, or harm-reduction guidance may still be provided. \\
5 & It is not possible to answer this safely as written. Consider reframing the question around a legitimate and non-harmful goal. \\
6 & The model will never assist with guidance that could put people or systems at risk. Safer alternatives and protective safeguards may be offered instead. \\
7 & Safety rules prohibit assistance with the proposed activity. General information about risks, detection, or mitigation is still available. \\
8 & I must decline to provide operational details for this activity. A non-actionable overview may be appropriate. \\
9 & Such instructions fall outside what the assistant is authorized to provide. I can support lawful and responsible alternatives. \\
10 & No actionable instructions will be provided for carrying out the proposed harm. I can redirect the discussion toward safety planning or recovery. \\
\bottomrule
\end{tabularx}
\end{table}
\end{minipage}
\par
\begin{multicols}{2}
\raggedcolumns

On 159 held-out harmful prompts with 30 samples each, pattern match is 96.10\% and the refusal-like rate is 97.38\%. Pattern frequencies are $[10.10,\allowbreak8.93,\allowbreak9.96,\allowbreak8.24,\allowbreak9.79,\allowbreak9.06,\allowbreak10.19,\allowbreak10.59,\allowbreak9.29,\allowbreak9.96]\%$ (total variation 0.0478 unconditionally and 0.0305 conditional on a match), so the ten surface forms are approximately balanced. Nevertheless, eight of the nine unique first tokens---covering nine patterns---select the same $-5$/L12 direction. Only \texttt{The} selects $-3$/L19, with cosine 0.0746 to the shared direction.

\end{multicols}
\par\noindent\begin{minipage}{\textwidth}
\begin{table}[H]
\caption{Refusal and target-pattern metrics before and after direction ablation.}
\label{tab:multi-ablation}
\centering
\normalsize
\begin{adjustbox}{max width=\linewidth}
\begin{tabular}{lccc}
\toprule
Condition & RR & 10-pattern match & RR change \\
\midrule
Baseline & 98.67 & 97.11 & -- \\
Shared $-5$/L12 ablated & 6.82 & 2.89 & $-91.85$ pp \\
Fresh \texttt{The} $-3$/L19 ablated & 97.71 & 92.84 & $-0.96$ pp \\
\bottomrule
\end{tabular}
\end{adjustbox}
\end{table}
\end{minipage}
\par
\begin{multicols}{2}
\raggedcolumns

Thus output diversity does not imply independent refusal mechanisms: ablating the shared direction collapses both refusal and pattern match, whereas ablating the approximately orthogonal \texttt{The} direction mainly changes surface form.

\end{multicols}
\par\noindent\begin{minipage}{\textwidth}
\subsubsection{Refusal-token probes after SafeTrain}
\label{app:per-round-results}

We re-extract a rank-one direction from the final three-round SafeTrain checkpoint using five first-output-token probes. Only \texttt{I} is a high-probability active prefix; the other four begin with very low baseline log-odds, so unchanged RR for those probes is not evidence of four independent safety mechanisms.

\begin{table}[H]
\caption{Fresh rank-one probes on the final SafeTrain model (32 prompts, 10 samples each).}
\label{tab:fresh-tokens}
\centering
\normalsize
\begin{adjustbox}{max width=\linewidth}
\begin{tabular}{lcccc}
\toprule
Probe $R$ & Selected source & Base RR & Ablated RR & Log-odds change \\
\midrule
$\{\texttt{I}\}$ & $-5$/L11 & 100.00 & 97.19 & $10.689\rightarrow7.385$ \\
$\{\texttt{Sorry}\}$ & $-1$/L4 & 100.00 & 100.00 & $-15.939\rightarrow-16.492$ \\
$\{\texttt{As}\}$ & $-1$/L4 & 100.00 & 100.00 & $-15.662\rightarrow-16.201$ \\
$\{\texttt{No}\}$ & $-4$/L21 & 100.00 & 100.00 & $-13.230\rightarrow-13.932$ \\
$\{\texttt{You}\}$ & $-4$/L18 & 100.00 & 100.00 & $-15.117\rightarrow-15.963$ \\
\bottomrule
\end{tabular}
\end{adjustbox}
\end{table}
\end{minipage}
\par
\begin{multicols}{2}
\raggedcolumns

\texttt{Sorry} and \texttt{As} select the same direction (cosine 1.0000), while \texttt{No} and \texttt{You} select distinct but similar directions (cosine 0.8446). The fresh final-model \texttt{I} direction and the initial-SFT $d_1$ have cosine 0.79198. The directions therefore remain strongly related after SafeTrain, although they are not identical. This diagnostic uses a 32$\times$10 stochastic cohort and is not numerically interchangeable with the matched 125/128-prompt greedy RSA evaluation.

\end{multicols}
\par\noindent\begin{minipage}{\textwidth}
\begin{table}[H]
\centering
\small
\caption{Candidate-pool control on frozen ReSAT checkpoints. The new pool contains disjoint prompts from PKU-SafeRLHF and Alpaca instructions. RSA HarmBench ASR is reported in \% (lower is better).}
\label{tab:w1-pool-control}
\begin{adjustbox}{max width=\linewidth}
\begin{tabular}{lcc}
\toprule
Backbone & Original pool & New pool \\
\midrule
Llama-3-8B   & 11.20  & \textbf{1.60 } \\
Mistral-7B   & 19.20  & \textbf{2.40 } \\
Qwen2.5-14B  & \textbf{8.80 } & 10.80  \\
\bottomrule
\end{tabular}
\end{adjustbox}
\end{table}
\end{minipage}
\par
\begin{multicols}{2}
\raggedcolumns

\end{multicols}
\par\noindent\begin{minipage}{\textwidth}
\begin{table}[H]
\centering
\small
\caption{Probe comparison on the original candidate pool (HarmBench ASR in \%, lower is better). \textsc{N/A} indicates that no candidate passed the predeclared eligibility gate.}
\label{tab:w1-probe-control}
\begin{adjustbox}{max width=\linewidth}
\begin{tabular}{lccc}
\toprule
Backbone & \texttt{I} & \texttt{I+Sorry} & \texttt{Sorry}-only \\
\midrule
Llama-3-8B   & 11.20  & 11.20  & \textsc{N/A} \\
Mistral-7B   & 19.20 & 19.20  & \textsc{N/A} \\
Qwen2.5-14B  & 8.80  & 8.80   & \textsc{N/A} \\
\bottomrule
\end{tabular}
\end{adjustbox}
\end{table}
\end{minipage}
\par
\begin{multicols}{2}
\raggedcolumns

\section{Unified evaluation protocol and cohorts}
\label{app:unified-results}
\label{app:main-table-protocol}

\paragraph{Metrics.}
Table~\ref{tab:main-jailbreak-utility-results} uses the mixed jailbreak scores defined in Section~\ref{sec:experiment-setup}, whereas Table~\ref{tab:main-three-family-results} reports independent HarmBench ASR and intervention-matched MMLU. Full-test MMLU/GSM8K and XSTest measure unablated capability and helpfulness. Lexical RR and its scoring rules are reported in Appendix~\ref{app:lexical-refusal-rates}; ASR is average of 3 judge classifer model.

\paragraph{Evaluation samples.}
Refusal-ablation evaluations use 150 prompts per run. Clean is the unablated baseline for the recorded RSA evaluation; Geometry/SOM retain their corresponding attack-run baselines. Evaluation directions are selected after training, separately from the SafeTrain banks. Saved responses are independently judged with the pinned HarmBench protocol and other classifier.
\paragraph{Attack endpoints and status.}
Single is self-seeded rank one; RSA reports the last accepted cumulative rank, not a test-selected optimum. If only rank one is accepted, RSA reuses Single, denoted $(1)$ where shown. The default gate is KL $\leq0.1$ with the recorded eligibility checks; Geometry reports rank five and SOM its validation-selected ordered six-direction endpoint. 

\end{multicols}
\par\noindent\begin{minipage}{\textwidth}
\section{Computational cost}
\label{app:computational-cost}

\subsection{Attack selection and defense training across model families}
\label{app:cost-ours}

Table~\ref{tab:computational-cost} gives single-run wall-clock times for Ours(F) in minutes. Runs use NVIDIA GH200 120GB GPUs: one for RSA and two for training, including the PAP generator. These are elapsed times, not aggregate GPU time; run-to-run variability was not measured.

\begin{table}[H]
\centering
\caption{Cost of Ours(F). SafeTrain uses 3, 2, and 1 rounds for Llama, Mistral, and Qwen, respectively; Mistral retains the same direction in round 2.}
\label{tab:computational-cost}
\normalsize
\setlength{\tabcolsep}{6pt}
\renewcommand{\arraystretch}{1.08}
\begin{adjustbox}{max width=\linewidth}
\begin{tabular}{@{}lcccc@{}}
\toprule
Model & \shortstack{RSA selection\\(min)} & \shortstack{SFT\\(min)} & \shortstack{SafeTrain\\(min)} & \shortstack{Full training\\pipeline (min)} \\
\midrule
Llama-3-8B & 10.3 & 154.2 & 387.6 & 619.2 \\
Mistral-7B & 6.1 & 154.8 & 206.4 & 396.0 \\
Qwen2.5-14B & 19.1 & 211.8 & 117.6 & 372.0 \\
\bottomrule
\end{tabular}
\end{adjustbox}
\end{table}
\end{minipage}
\par
\begin{multicols}{2}
\raggedcolumns

\noindent\textbf{Attack timing.} RSA selection sums four self-seeded candidate-scoring sweeps on the final checkpoint, including unsuccessful fourth sweeps for Mistral and Qwen, rounded to 0.1 minute. Loading, activation extraction, and downstream evaluation are excluded.

\noindent\textbf{Training timing.} SFT and SafeTrain include online attacks, optimization, validation, and checkpoint saving. Full-pipeline totals additionally include loading, direction search, smoke checks, and in-chain diagnostics. Only successful jobs are summed; queueing, failed runs, hyperparameter search, offline data generation, and final external benchmarks are excluded.

\end{multicols}
\par\noindent\begin{minipage}{\textwidth}
\subsection{Comparison with multi-direction ablation methods}
\label{app:cost-multidirection}

Table~\ref{tab:cost-multidirection} compares our GH200 measurements of RSA, Geometry \citep{wollschlager2025geometry}, and SOM \citep{piras2026som} on the same seven Llama-family checkpoints. RSA uses one GPU; Geometry and SOM allocate two for the target and judge. Search budgets, probes, and timing boundaries differ. Defense training and final held-out evaluation are excluded.

\begin{table}[H]
\centering
\caption{Core attack-search wall-clock time on seven Llama-family checkpoints (minutes).}
\label{tab:cost-multidirection}
\normalsize
\setlength{\tabcolsep}{7pt}
\renewcommand{\arraystretch}{1.12}
\begin{adjustbox}{max width=\linewidth}
\begin{tabular}{@{}lccc@{}}
\toprule
Target & \shortstack{RSA selection\\(min)} & \shortstack{Geometry learning\\+ selection (min)} & \shortstack{SOM search\\(min)} \\
\midrule
Base & 4.00 & 101.35 & 1310.37 \\
ReFAT & 6.70 & 93.20 & 1128.52 \\
DeepRefusal & 5.10 & 100.67 & 285.48 \\
Ours(ReSAT) & 10.30 & 76.30 & 111.33 \\
MixAT & 9.30 & 71.02 & 969.40 \\
LAT & 4.90 & 60.50 & 1082.12$^{\dagger}$ \\
CAT & 4.70 & 61.48 & 888.52 \\
\bottomrule
\end{tabular}
\end{adjustbox}

\smallskip
\begin{minipage}{0.96\linewidth}
\normalsize
\end{minipage}
\end{table}
\end{minipage}
\par
\begin{multicols}{2}
\raggedcolumns

\noindent\textbf{RSA search.} Each of up to four self-seeded sweeps scores 160 candidates. Base and ReFAT stop at rank two after three sweeps; others reach rank four. Timers count unsuccessful sweeps and Base's reused first-direction search. MixAT/CAT use \texttt{Sorry}. Loading, activation extraction, and downstream generation/evaluation are excluded.

\noindent\textbf{Geometry search.} Five directions are learned sequentially, with one initialization and a two-epoch cap per direction. Timing reconstructed from target-audit and saved-direction timestamps includes data preparation, optimization, and validation, excluding initial loading and downstream generation/evaluation (including judging). MixAT/CAT use \texttt{Sorry}.

\noindent\textbf{SOM search.} We run 512 Bayesian-optimization trials over six directions using 41 validation prompts. Timing includes per-trial loading, generation, internal judging, and bookkeeping, excluding initial representation extraction/direction construction and final held-out evaluation.


\end{multicols}
\par\noindent\begin{minipage}{\textwidth}
\section{Lexical Refusal Rates under Ablation}
\label{app:lexical-refusal-rates}

\noindent\textbf{Lexical scoring.}Clean/Single/RSA use the recorded union of 12 case-insensitive refusal substrings and broader refusal-language regular expressions; specified v2 rows also recognize whole-response short refusals. Geometry/SOM retain their original 12-substring scorer. Table~\ref{tab:ablation-lexical-rr} reports lexical RR for the Table~\ref{tab:main-three-family-results} checkpoints with available RR evaluations. Attack endpoints and status definitions follow Appendix~\ref{app:main-table-protocol}.

\begingroup
\begin{table}[H]
\centering
\caption{Lexical refusal rates (RR, \%; higher is better) for the Table~\ref{tab:main-three-family-results} checkpoints with available RR evaluations.}
\label{tab:ablation-lexical-rr}
\normalsize
\setlength{\tabcolsep}{3pt}
\renewcommand{\arraystretch}{1.08}
\begin{adjustbox}{max width=\linewidth}
\begin{tabular}{@{}lccccc@{}}
\toprule
Model & Clean & Single & Geometry & SOM & RSA (Ours) \\
\midrule
\multicolumn{6}{@{}l}{\textbf{Llama-3-8B}} \\
Base & 97.66 & 4.69 & 0.00 & 11.32 & 3.12$^{(2)}$ \\
LAT & \textbf{100.00} & 11.72 & 0.00 & 11.95 & 4.69$^{(4)}$ \\
CAT & \textbf{100.00} & 75.00 & \underline{42.14} & 8.81 & 48.44$^{(4)}$ \\
MixAT & \textbf{100.00} & 14.06 & 0.00 & 4.40 & 6.25$^{(4)}$ \\
ReFAT & \underline{99.22} & 3.12 & 0.63 & 4.40 & 2.34$^{(2)}$ \\
DeepRefusal & \textbf{100.00} & \textbf{100.00} & \textbf{100.00} & \underline{71.70} & \underline{73.44}$^{(4)}$ \\
\textbf{Ours(H)} & \textbf{100.00} & \underline{95.20} & \textbf{100.00} & \textbf{100.00} & \textbf{84.80}$^{(4)}$ \\
\midrule
\multicolumn{6}{@{}l}{\textbf{Mistral-7B}} \\
Base & 48.44 & 5.47 & 48.44 & 3.14 & 5.47$^{(1)}$ \\
LAT & 92.19 & 4.69 & \underline{76.10} & 2.52 & 6.25$^{(2)}$ \\
CAT & 97.66 & \underline{64.84} & 2.52 & 18.24 & 6.25$^{(4)}$ \\
MixAT & \underline{99.22} & 53.91 & 33.96 & 0.63 & 14.84$^{(3)}$ \\
ReFAT & \textbf{100.00} & 42.97 & 3.77 & \underline{35.85} & \underline{42.97}$^{(1)}$ \\
DeepRefusal & 89.84 & 2.34 & 47.80 & 0.00 & 2.34$^{(1)}$ \\
\textbf{Ours(H-R12)} & \textbf{100.00} & \textbf{97.60} & \textbf{100.00} & \textbf{62.34}& \textbf{77.60}$^{(3)}$ \\
\midrule
\multicolumn{6}{@{}l}{\textbf{Qwen2.5-14B}} \\
Base & 81.25 & 4.69 & 0.00 & 2.52 & 4.69$^{(1)}$ \\
LAT & \textbf{100.00} & 13.28 & 1.26 & 1.89 & 13.28$^{(1)}$ \\
CAT & \underline{99.22} & \underline{98.44} & \underline{7.55} & \underline{96.23} & \textbf{98.44}$^{(1)}$ \\
MixAT & \textbf{100.00} & \textbf{99.22} & 0.00 & \textbf{100.00} & 72.66$^{(3)}$ \\
ReFAT & \underline{99.22} & 62.50 & 0.00 & 1.26 & 32.03$^{(2)}$ \\
DeepRefusal & \underline{99.22} & 28.91 & 0.00 & 7.55 & 3.12$^{(3)}$ \\
\textbf{Ours(MB-v2)} & \textbf{100.00} & 98.40 & \textbf{92.21} & 74.68 & \underline{88.00}$^{(2)}$ \\

\bottomrule
\end{tabular}
\end{adjustbox}
\end{table}
\endgroup
\end{minipage}
\par
\begin{multicols}{2}
\raggedcolumns

\end{multicols}
\par\noindent\begin{minipage}{\textwidth}
\section{Ablation Study Details}
\label{app:llama-ablation-supplement}
\label{app:ablation-study-details}

Table~\ref{tab:ablation-training-setups} summarizes the six Llama-3-8B ablation groups in Table~\ref{tab:training-recipe-ablation}, listing training settings and sampling shares without evaluation results. All setups use the same instruction-tuned backbone and the implementation in Appendix~\ref{app:training-details}. The reference configuration is denoted \textbf{Ours} throughout this appendix.

\begin{table}[H]
\centering
\caption{Training setups and sampling shares for the six ablation groups. All shares are percentages; Rds. counts completed SafeTrain rounds.}
\label{tab:ablation-training-setups}
\normalsize
\setlength{\tabcolsep}{3pt}
\begin{adjustbox}{max width=\linewidth}
\begin{tabular}{@{}lccc@{}}
\toprule
Setup & SFT shares & SafeTrain shares & Rds. \\
 & C/A/P/U/HB & C/R/A/P/U/HB & \\
\midrule
\multicolumn{4}{@{}l}{\textit{(i) SFT initialization}} \\
Original SFT & 50/0/0/50/0 & 25/12.5/6.25/6.25/45/5 & 3 \\
Mixed SFT & 20/5/5/50/20 & 25/12.5/6.25/6.25/45/5 & 3 \\
No initial SFT & --- & 25/12.5/6.25/6.25/45/5 & 3 \\
\midrule
\multicolumn{4}{@{}l}{\textit{(ii) Input-attack training}} \\
No input attacks & 30/0/0/50/20 & 10/35/0/0/35/20 & 3 \\
With input attacks (Ours) & 20/5/5/50/20 & 10/22.5/6.25/6.25/35/20 & 3 \\
\midrule
\multicolumn{4}{@{}l}{\textit{(iii) Hard-benign supervision}} \\
No Hard Benign & 20/5/5/70/0 & 10/22.5/6.25/6.25/55/0 & 3 \\
With Hard Benign (Ours) & 20/5/5/50/20 & 10/22.5/6.25/6.25/35/20 & 3 \\
\midrule
\multicolumn{4}{@{}l}{\textit{(iv) Training rounds}} \\
Ours after round 1 & 20/5/5/50/20 & 10/22.5/6.25/6.25/35/20 & 1 \\
Ours after round 2 & 20/5/5/50/20 & 10/22.5/6.25/6.25/35/20 & 2 \\
Ours after round 3 & 20/5/5/50/20 & 10/22.5/6.25/6.25/35/20 & 3 \\
\midrule
\multicolumn{4}{@{}l}{\textit{(v) Complementary directions}} \\
Repeated $d_1$ & 20/5/5/50/20 & 10/22.5/6.25/6.25/35/20 & 3 \\
Cumulative directions (Ours) & 20/5/5/50/20 & 10/22.5/6.25/6.25/35/20 & 3 \\
\midrule
\multicolumn{4}{@{}l}{\textit{(vi) Training-time RSA}} \\
Without training RSA$^{\dagger}$ & 20/5/5/50/20 & 10/22.5/6.25/6.25/35/20 & 3 \\
With training RSA (Ours) & 20/5/5/50/20 & 10/22.5/6.25/6.25/35/20 & 3 \\
\bottomrule
\end{tabular}
\end{adjustbox}
\end{table}
\end{minipage}
\par
\begin{multicols}{2}
\raggedcolumns

\paragraph{Mixture definitions and budget.}
C denotes clean harmful-prompt refusal; R, refusal under direction ablation; A, continuous input attacks; P, PAP rewriting followed by continuous attacks; U, ordinary utility; and HB, hard-benign supervision. Harmful branches supervise safe refusals, whereas U and HB supervise helpful answers. SFT never uses direction ablation; SafeTrain applies it to R/A/P except in the no-RSA control. SFT runs for four epochs where present. Each SafeTrain round contains 4,000 presentations and 32 optimizer updates, giving 32/64/96 updates after 1/2/3 rounds.

\paragraph{Controlled changes.}
Group (i) holds the SafeTrain mixture and three-round schedule fixed, but does not match total training compute. In (ii), removing input attacks reallocates their SFT shares to C and SafeTrain shares to R; source queries, safe targets, sample order, microbatch membership, hard-benign exposure, and update budgets are preserved. Group (iii) replaces HB with U in both stages, preserving the harmful branches, total sample count, update budget, and benign share. Group (iv) uses successive checkpoints from one run; round 3 is the predefined final model. In (v), Repeated $d_1$ reuses the same first-round checkpoint, then fixes that direction for two further rounds, giving training-bank ranks $(1,1,1)$ instead of Ours' $(1,2,3)$; both use 96 updates. Final-model RSA searches fresh directions independently of the training banks.

\end{multicols}
\par\noindent\begin{minipage}{\textwidth}
\noindent $^{\dagger}$In (vi), both setups start from the same mixed-SFT checkpoint. Without training RSA retains Ours' sample quotas, source order, safe targets, continuous/PAP attacks, and 96-update budget, but disables direction ablation in both inner attack search and outer training. R/A/P remain sampling categories with their projections removed. This control matches optimizer updates, not FLOPs or the realized adversarial inputs.
\end{minipage}
\par
\begin{multicols}{2}
\raggedcolumns


\end{multicols}
\par\noindent\begin{minipage}{\textwidth}
\section{Cumulative Direction-History Mechanism Analysis}
\label{app:cumulative-mechanism}

Figure~\ref{fig:cumulative-mechanism} reports the complete set of diagnostics accompanying Section~\ref{sec:mechanism-analysis}.

\begin{figure}[H]\centering\includegraphics[width=0.85\textwidth]{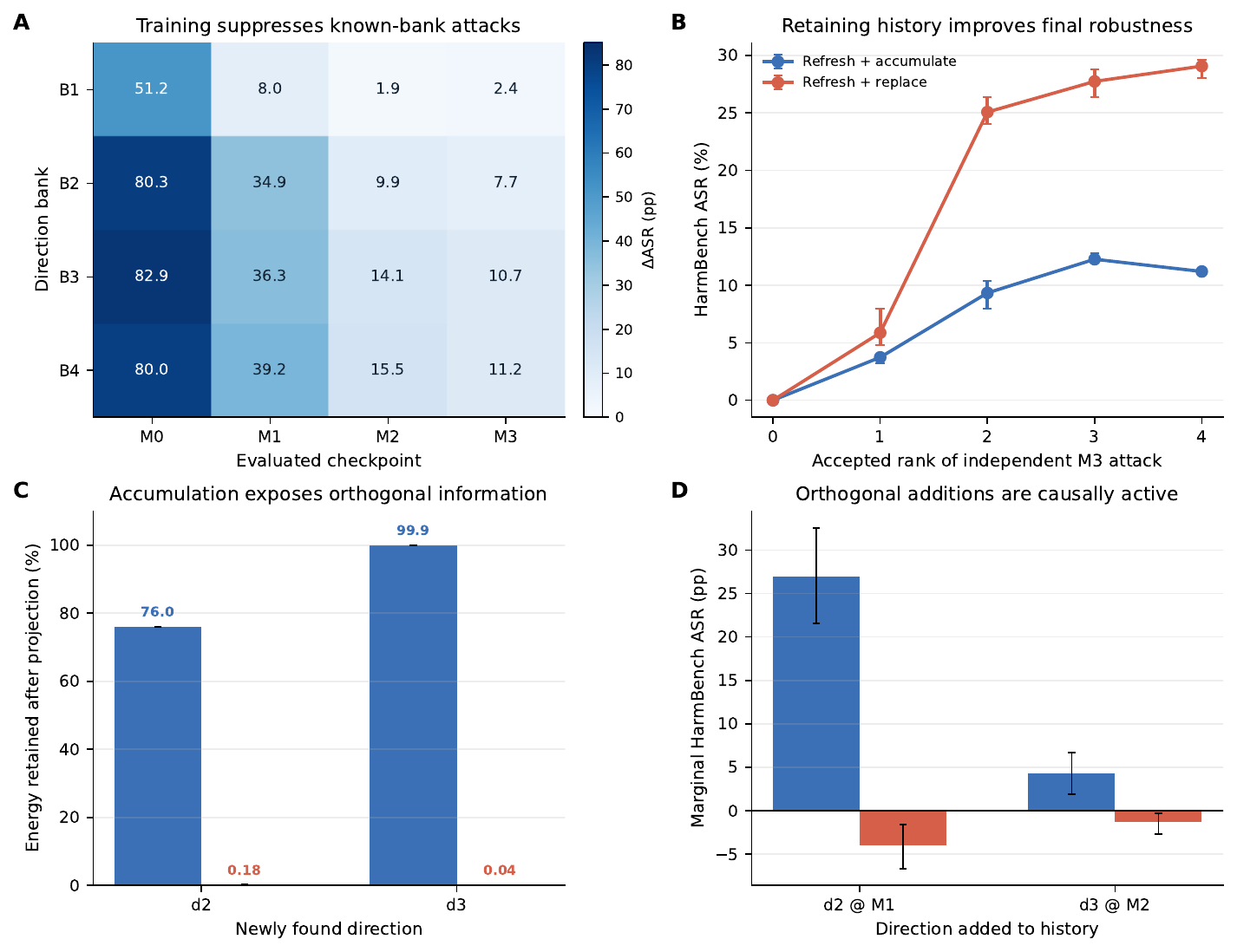}
    \caption{\textbf{Mechanism analysis of cumulative direction history on Llama-3-8B.} \textbf{(A)} SafeTrain progressively suppresses attacks from known direction banks. \textbf{(B)} Retaining history yields lower independently recoverable $M_3$ ASR than replacing the bank. \textbf{(C)} Accumulation exposes information outside the historical span, whereas replacement largely rediscovers that span. \textbf{(D)} The complementary directions found under accumulation have positive marginal causal effects. Results use three matched seeds and 125 held-out HarmBench prompts per seed. Error bars in Panels B and C show the seed range; error bars in Panel D show paired hierarchical-bootstrap 95\% confidence intervals over seeds and prompts.}
    \label{fig:cumulative-mechanism}\end{figure}
\end{minipage}
\par
\begin{multicols}{2}
\raggedcolumns

\noindent\textbf{Shared protocol.} We use the Llama-3-8B checkpoints $M_0,\ldots,M_3$, where $M_0$ is the SFT-initialized model and $M_t$ is saved after SafeTrain round $t$. Unless stated otherwise, all results aggregate three matched seeds and use the same 125 held-out HarmBench prompts per seed. ASR is scored by the independent HarmBench classifier; no directions are selected on this held-out evaluation set. For schedule $s$, let $B^{(s)}_{<j}=\operatorname{span}(d^{(s)}_1,\ldots,d^{(s)}_{j-1})$ denote the schedule-specific prefix. Refresh+accumulate supplies this prefix during direction discovery and training, whereas refresh+replace discards it; for replacement, we reconstruct the prefix only post hoc for the matched diagnostics in Panels C and D.

\noindent\textbf{Panel A: suppression of known bank prefixes.} \textbf{Setup.} For every checkpoint $M_t$, we replay ablation of the cumulative training-bank prefixes $B_k=\operatorname{span}(d_1,\ldots,d_k)$, $k\in\{1,2,3,4\}$, without re-estimating the directions, and report the ASR increase relative to the corresponding clean checkpoint. \textbf{Conclusion.} SafeTrain progressively defends against the interventions already included in the bank: the $B_1$ effect falls from 51.2 pp on $M_0$ to 8.0, 1.9, and 2.4 pp on $M_1$, $M_2$, and $M_3$, while the $B_2$--$B_4$ effects likewise fall from 80.3--82.9 pp on $M_0$ to 7.7--11.2 pp on $M_3$. Thus, later checkpoints acquire robustness to the known bank rather than merely changing their clean refusal rate.

\noindent\textbf{Panel B: refresh-and-accumulate versus refresh-and-replace.} \textbf{Setup.} We compare two matched SafeTrain schedules with the same initialization, data quotas, update budget, and refresh times. Refresh+accumulate retains every previously found direction, whereas refresh+replace discards the old bank and trains only against the direction re-estimated in the current round. We then attack each final $M_3$ checkpoint independently and report HarmBench ASR at accepted attack ranks zero through four. \textbf{Conclusion.} At rank four, replacement reaches 29.07\% ASR, compared with 11.20\% under accumulation, a paired increase of 17.87 pp (95\% CI: $[14.13,21.87]$). Retaining direction history is therefore necessary for the final robustness gain.

\noindent\textbf{Panel C: orthogonal information in refreshed directions.} \textbf{Setup.} At the second and third refreshes, we project the newly found candidates $d_2$ and $d_3$ onto the orthogonal complement of the preceding historical span $B_{<j}$ and measure the retained squared-energy ratio $100\lVert P_{B_{<j}^{\perp}}d_j\rVert_2^2/\lVert d_j\rVert_2^2$. The same calculation is applied to candidates from both schedules. \textbf{Conclusion.} Under accumulation, $d_2$ and $d_3$ retain 76.0\% and 99.9\% of their energy after projection; replacement retains only 0.185\% and 0.043\%. Accumulation therefore exposes directions that are genuinely complementary to the covered span, whereas replacement almost entirely rediscovers previously covered directions.

\noindent\textbf{Panel D: causal validation of the orthogonal information.} \textbf{Setup.} For each refreshed candidate, we measure its held-out marginal causal effect by adding it to the preceding bank and computing $\mathrm{ASR}(B_{<j}\cup\{d_j\})-\mathrm{ASR}(B_{<j})$. We evaluate accumulated and replacement candidates with the same prompts and classifier. \textbf{Conclusion.} The accumulated $d_2$ and $d_3$ increase ASR by 26.93 and 4.27 pp, respectively, whereas the replacement candidates yield no positive marginal gain. The orthogonal information identified in Panel C is therefore causally active refusal vulnerability, not an incidental representation rotation.

\end{multicols}
\par\noindent\begin{minipage}{\textwidth}
\section{Results with Native Thinking Enabled}
\label{app:native-thinking-results}

\begin{table}[H]
\centering
\caption{RSA on Qwen3-8B with native thinking enabled, before ReSAT training.}
\label{tab:zmzx-w2-thinking-rsa}
\begingroup
\fontsize{9}{11}\selectfont
\setlength{\tabcolsep}{3pt}
\renewcommand{\arraystretch}{1.15}
\begin{tabular*}{\linewidth}{@{\extracolsep{\fill}}lccc@{}}
\toprule
Condition & HarmBench ASR $\uparrow$ & Lexical RR $\uparrow$ & MMLU accuracy \\
\midrule
Clean thinking & 9.6\%  & 68.8\% & 79.75\% \\
    RSA thinking(ours) & \textbf{88.8\% } & \textbf{14.4\%} & 78.25\% \\
\bottomrule
\end{tabular*}
\endgroup
\end{table}

\begin{table}[H]
\centering
\caption{ReSAT on Qwen3-8B with native reasoning enabled. Clean, Single, RSA (ours), and HumanJB report HarmBench ASR. XSTest reports benign compliance and MMLU reports accuracy.}
\label{tab:zmzx-w2-thinking-defense}
\begingroup
\fontsize{9}{11}\selectfont
\setlength{\tabcolsep}{3pt}
\renewcommand{\arraystretch}{1.15}
\begin{tabular*}{\linewidth}{@{\extracolsep{\fill}}lcccccc@{}}
\toprule
Model & Clean $\downarrow$ & Single $\downarrow$ & RSA $\downarrow$ & XSTest $\uparrow$ & MMLU $\uparrow$ & HumanJB $\downarrow$ \\
\midrule
ReSAT Qwen3-8B (reasoning) & 11.20\% & 8.00\% & 8.00\% & 100.00\% & 81.82\% & 14.81\% \\
\bottomrule
\end{tabular*}
\endgroup
\end{table}

\section{Additional Results on Gemma-7B-it}
\label{app:gemma-additional-results}

\begin{table}[H]
\centering
\caption{Gemma-7B-it: refusal-ablation attacks, jailbreaks, and utility. ASR is measured by HarmBench}
\label{tab:gyhm-w3-gemma-results}
\label{tab:gyhm-w3-gemma-utility}
\label{tab:gyhm-w3-gemma-rsa}
\begingroup
\fontsize{9}{11}\selectfont
\setlength{\tabcolsep}{3pt}
\renewcommand{\arraystretch}{1.15}
\begin{tabular*}{\linewidth}{@{\extracolsep{\fill}}lccccccc@{}}
\toprule
& \multicolumn{3}{c}{\shortstack{Attack ASR$\downarrow$ / MMLU$\uparrow$}} & \multicolumn{2}{c}{Jailbreak ASR$\downarrow$} & \multicolumn{2}{c}{Utility} \\
\cmidrule(lr){2-4}\cmidrule(lr){5-6}\cmidrule(l){7-8}
Model / Gemma-7B & Clean & Single & RSA 
& HumanJB
& GCG
& MMLU-400$\uparrow$
& XSTest$\uparrow$ \\
\midrule

Base & 3.20 / 46.75 & 88.80 / 47.25 & 88.80 / 47.25  & \shortstack{\\6.6} & \shortstack{32} & \shortstack{46.75} & 92 \\
CAT & 2.40 / 53.75 & 36.00 / 53.25 & 63.20 / 52.00  & \shortstack{\\5.2} & 28 & \shortstack{54.50} & 19 \\
LAT & 0.80 / 46.25 & 71.20 / 47.75 & 72.00 / 47.25  & \shortstack{0.8} & 22 & \shortstack{46.50} & 64 \\
DeepRefusal & 0.00 / 44.00 & 46.40 / 43.75 & 46.40 / 43.75 & \shortstack{0.8} & 14 & \shortstack{43.25} & 78 \\
ReSAT (Ours) & 0.00 / 49.25 & 9.60 / 50.75 & 7.20 / 49.50  & \shortstack{0.0} & \shortstack{2} & \shortstack{49.50} & 41 \\

\bottomrule
\end{tabular*}
\endgroup
\end{table}

\end{minipage}
\par
\begin{multicols}{2}
\raggedcolumns


\end{multicols}
\par\noindent\begin{minipage}{\textwidth}
\section{KL ablation}
\label{app:kl-ablation}

The KL we use for ReSAT training is 0.2 and RSA ablation attack is 0.1, but We also vary the RSA selection threshold to $\mathrm{KL}\leq0.2$ and $\mathrm{KL}\leq0.5$ on seven fixed checkpoints from each of three model families. Directions are freshly extracted from each fixed target and expanded adaptively. The KL gate measures benign first-token distribution divergence during selection, not an upper bound on MMLU degradation. Table~\ref{tab:kl-ablation-endpoints} reports the \emph{last accepted cumulative rank}, not the rank with the highest test ASR. A rank below four means that no next candidate passed the complete selection constraints. ASR is the independent HarmBench classifier's positive-label count divided by the original cohort size. RR is lexical refusal rate, and MMLU is utility benchmark. 


\begin{table}[H]
\centering
\caption{RSA KL ablation (\%). \emph{$\mathrm{KL}\leq0.1$} reproduces the RSA ASR/MMLU columns of main paper Table~\ref{tab:main-three-family-results} and the corresponding lexical RR from Table~\ref{tab:ablation-lexical-rr}, under their original settings. In all three groups, $r$ is the last accepted cumulative rank; the first group uses the ranks corresponding to the main-table RSA endpoints. RR is lexical refusal rate, MMLU is utility accuracy.}
\label{tab:kl-ablation-endpoints}
\begingroup
\small
\setlength{\tabcolsep}{3.2pt}
\renewcommand{\arraystretch}{1.08}
\begin{adjustbox}{max width=\linewidth}
\begin{tabular}{@{}l*{12}{r}@{}}
\toprule
& \multicolumn{4}{c}{$\mathrm{KL}\leq0.1$} & \multicolumn{4}{c}{$\mathrm{KL}\leq0.2$} & \multicolumn{4}{c}{$\mathrm{KL}\leq0.5$} \\
\cmidrule(lr){2-5}\cmidrule(lr){6-9}\cmidrule(lr){10-13}
Model & $r$ & ASR$\downarrow$ & RR$\uparrow$ & MMLU$\uparrow$ & $r$ & ASR$\downarrow$ & RR$\uparrow$ & MMLU$\uparrow$ & $r$ & ASR$\downarrow$ & RR$\uparrow$ & MMLU$\uparrow$ \\
\midrule
\multicolumn{13}{@{}l}{\textbf{Llama-3-8B}} \\
Base & 2 & 92.19 & 3.12 & 64.25 & 3 & 94.53 & 7.81 & 64.00 & 4 & 88.28 & 4.69 & 63.25 \\
LAT & 4 & 86.72 & 4.69 & 62.75 & 3 & 92.19 & 4.69 & 63.50 & 4 & 88.28 & 5.47 & 64.25 \\
CAT & 4 & 45.31 & 48.44 & 64.00 & 4 & 58.59 & 32.81 & 64.00 & 4 & 74.22 & 15.63 & 65.75 \\
MixAT & 4 & 82.03 & 6.25 & 60.25 & 4 & 64.84 & 24.22 & 61.25 & 4 & 79.69 & 6.25 & 60.25 \\
ReFAT & 2 & 92.19 & 2.34 & 65.25 & 3 & 84.38 & 2.34 & 64.50 & 3 & 88.28 & 2.34 & 65.50 \\
DeepRefusal & 4 & 23.44 & 73.44 & 64.25 & 4 & 4.69 & 93.75 & 65.00 & 4 & 3.13 & 95.31 & 65.25 \\
\textbf{Ours (ReSAT)} & 4 & 11.20 & 84.80 & 64.75 & 4 & 16.80 & 78.40 & 64.50 & 4 & 16.80 & 78.40 & 64.50 \\
\midrule
\multicolumn{13}{@{}l}{\textbf{Mistral-7B}} \\
Base & 1 & 69.53 & 5.47 & 59.50 & 1 & 73.44 & 3.13 & 56.25 & 1 & 72.66 & 3.91 & 56.25 \\
LAT & 2 & 81.25 & 6.25 & 54.00 & 2 & 81.25 & 5.47 & 54.00 & 2 & 81.25 & 6.25 & 54.00 \\
CAT & 4 & 75.78 & 6.25 & 53.25 & 4 & 75.78 & 6.25 & 53.25 & 4 & 75.78 & 5.47 & 53.25 \\
MixAT & 3 & 65.62 & 14.84 & 55.75 & 4 & 71.09 & 10.94 & 50.75 & 4 & 71.09 & 10.94 & 50.75 \\
ReFAT & 1 & 28.91 & 42.97 & 57.25 & 4 & 71.09 & 3.91 & 54.75 & 4 & 59.38 & 2.34 & 42.25 \\
DeepRefusal & 1 & 78.91 & 2.34 & 57.75 & 2 & 42.97 & 3.91 & 53.50 & 2 & 43.75 & 3.91 & 53.50 \\
\textbf{Ours (ReSAT)} & 3 & 20.00 & 77.60 & 57.25 & 4 & 16.00 & 83.20 & 54.25 & 4 & 20.00 & 78.40 & 48.00 \\
\midrule
\multicolumn{13}{@{}l}{\textbf{Qwen2.5-14B}} \\
Base & 1 & 79.69 & 4.69 & 76.50 & 1 & 81.25 & 4.69 & 76.50 & 1 & 78.91 & 1.56 & 75.50 \\
LAT & 1 & 47.66 & 13.28 & 76.00 & 2 & 71.09 & 0.00 & 76.00 & 3 & 94.53 & 3.13 & 74.50 \\
CAT & 1 & 1.56 & 98.44 & 75.50 & 1 & 14.84 & 78.91 & 75.75 & 4 & 35.94 & 15.63 & 75.25 \\
MixAT & 3 & 29.69 & 72.66 & 70.25 & 1 & 9.38 & 88.28 & 77.25 & 3 & 31.25 & 72.66 & 70.25 \\
ReFAT & 2 & 64.06 & 32.03 & 75.50 & 2 & 56.25 & 32.81 & 77.25 & 2 & 73.44 & 0.78 & 76.75 \\
DeepRefusal & 3 & 72.66 & 3.12 & 75.75 & 3 & 91.41 & 6.25 & 75.25 & 4 & 79.69 & 5.47 & 72.50 \\
\textbf{Ours (ReSAT)} & 2 & 8.80 & 88.00 & 79.00 & 3 & 19.20 & 76.00 & 77.75 & 4 & 69.60 & 12.00 & 73.25 \\
\bottomrule
\end{tabular}
\end{adjustbox}
\endgroup
\end{table}
\end{minipage}
\par
\begin{multicols}{2}
\raggedcolumns




\end{multicols}
\par\noindent\begin{minipage}{\textwidth}
\section{Seed Reproducibility Across Model Families}
\label{app:seed-reproducibility}

This appendix reports every completed training-seed evaluation for the primary
Llama-3-8B Ours, Mistral-7B Ours, and Qwen2.5-14B Ours
recipes.  Seeds 20260908, 20260909, and 20260910 are distinct trained
checkpoints for each family.  Each refusal-ablation cell is reported as
independent HarmBench ASR / lexical refusal rate (RR) / MMLU accuracy, in
percent.  


\begin{table}[H]
\centering
\caption{Complete refusal-ablation seed results (\%). Each entry is HarmBench
ASR / lexical RR / MMLU.  The parenthesized value is the last accepted RSA
rank.}
\label{tab:seed-refusal-ablation}
\scriptsize
\setlength{\tabcolsep}{2.5pt}
\resizebox{\textwidth}{!}{%
\begin{adjustbox}{max width=\linewidth}
\begin{tabular}{@{}llccccc@{}}
\toprule
Family & Seed & Clean & Single & Geometry & SOM & RSA endpoint \\
\midrule
\multirow{3}{*}{Llama-3-8B Ours}
 & 20260908 & 0.00/100.00/65.75 & 4.00/95.20/63.75 & 0.00/100.00/65.50 & 0.00/100.00/66.00 & 11.20/84.80/64.75 (4) \\
 & 20260909 & 0.00/100.00/66.00 & 4.00/95.20/63.50 & 0.00/100.00/65.25 & 0.00/100.00/66.50 & 11.20/84.80/65.00 (4) \\
 & 20260910 & 0.00/100.00/65.25 & 3.20/96.00/63.75 & 0.00/100.00/65.25& 0.00/100.00/66.75 & 11.20/85.60/65.00 (4) \\
\midrule
\multirow{3}{*}{Mistral-7B Ours}
 & 20260908 & 0.00/100.00/57.25 & 1.60/97.60/57.25 & 0.00/100.00/55.50 & 28.57/62.34/47.25 & 20.00/77.60/57.25 (3) \\
 & 20260909 & 0.00/100.00/57.25 & 1.60/97.60/57.75 & 0.00/100.00/54.25 & 35.71/62.34/47.50 & 20.00/76.80/57.25 (3) \\
 & 20260910 & 0.00/100.00/57.50 & 1.60/97.60/57.75 & 0.00/100.00/53.75 & 30.52/62.34/39.50 & 19.20/78.40/57.00 (3) \\
\midrule
\multirow{3}{*}{Qwen2.5-14B Ours}
 & 20260908 & 0.00/100.00/78.50 & 1.50/98.40/79.00 & 7.14/92.21/78.50 & 22.73/74.68/76.75 & 8.80/88.00/79.00 (2) \\
 & 20260909 & 0.00/100.00/79.00 & 1.50/98.40/79.00 & 20.02/16.88/78.50 & 26.62/71.43/76.50 & 8.80/88.80/79.00 (2) \\
 & 20260910 & 0.00/100.00/78.50 & 1.50/98.40/79.00 & 2.60/97.40/78.50 & 18.83/78.57/75.50 & 8.80/89.60/79.25 (2) \\
\bottomrule
\end{tabular}
\end{adjustbox}}
\end{table}
\end{minipage}
\par
\begin{multicols}{2}
\raggedcolumns

\end{multicols}
\par\noindent\begin{minipage}{\textwidth}
\begin{table}[H]
\centering
\caption{Complete jailbreak and utility results across the seed runs (\%).
NoAtk--RFA are the paper's composite safety scores (lower is better); MMLU,
GSM8K, and XSTest are utility scores (higher is better).}
\label{tab:seed-jailbreak-utility}
\scriptsize
\setlength{\tabcolsep}{2.8pt}
\resizebox{\textwidth}{!}{%
\begin{adjustbox}{max width=\linewidth}
\begin{tabular}{@{}llrrrrrrrrrrr@{}}
\toprule
Family & Seed & NoAtk & AutoDAN & HumanJB & GCG & PAIR & Inp. & Soft & RFA & MMLU & GSM8K & XSTest \\
\midrule
\multirow{3}{*}{Llama-3-8B}
 & 20260908 & 0.04 & 0.04 & 0.04 & 1.05 & 1.19 & 47.31 & 82.12 & 57.25 & 66.49 & 73.09 & 53.00 \\
 & 20260909 & 0.04 & 0.04 & 0.04 & 2.37 & 2.81 & 48.94 & 84.19 & 56.95 & 66.52 & 72.48 & 53.00 \\
 & 20260910 & 0.04 & 0.04 & 0.04 & 1.56 & 1.35 & 46.72 & 83.57 & 57.42 & 67.04 & 73.46 & 55.00 \\
\midrule
\multirow{3}{*}{Mistral-7B}
 & 20260908 & 0.04 & 4.12 & 0.07 & 53.83 & 30.49 & 64.63 & 80.51 & 47.29 & 59.22 & 42.99 & 57.00 \\
 & 20260909 & 0.04 & 5.48 & 0.07 & 57.48 & 38.10 & 64.26 & 81.10 & 47.93 & 59.19 & 42.46 & 60.00 \\
 & 20260910 & 0.04 & 7.22 & 0.07 & 61.86 & 33.02 & 63.44 & 81.44 & 47.69 & 59.19 & 42.61 & 56.00 \\
\midrule
\multirow{3}{*}{Qwen2.5-14B}
 & 20260908 & 0.04 & 4.58 & 0.65 & 52.48 & 44.21 & 80.60 & 83.75 & 52.07 & 78.93 & 86.28 & 47.00 \\
 & 20260909 & 0.04 & 3.85 & 0.66 & 54.11 & 50.69 & 79.47 & 85.29 & 53.49 & 78.86 & 86.50 & 49.00 \\
 & 20260910 & 0.04 & 0.04 & 0.61 & 45.96 & 48.28 & 80.21 & 84.44 & 52.84 & 78.89 & 86.35 & 49.00 \\
\bottomrule
\end{tabular}
\end{adjustbox}}
\end{table}
\end{minipage}
\par
\begin{multicols}{2}
\raggedcolumns


\end{multicols}
\end{document}